\documentclass[12pt]{article}
\begin{document}
\hyphenation{Schwarz-schild}
\hyphenation{Be-ken-stein}
\hyphenation{Haw-king}
\begin{flushright}
NSF-ITP-97-148 \\
hep-th/9712253  \\
\end{flushright}
\vskip1truein
\begin{center}
{\huge\bf The Bekenstein Formula \\ 
       and \\ 
       String Theory \\ }
{\Large\bf (N-brane Theory)}

\vskip0.5truein
        Amanda W. Peet \\ 
        Institute for Theoretical Physics \\
        University of California \\ 
        Santa Barbara \  CA \  93106-4030 U.S.A. \\
        {\tt peet@itp.ucsb.edu}  \\

\vskip0.5truein
{\bf Abstract} 
\end{center}
\noindent A review of recent progress in string theory concerning the
Bekenstein formula for black hole entropy is given.  Topics discussed
include $p$-branes, D-branes and supersymmetry; the correspondence
principle; the D- and M-brane approach to black hole entropy; the
D-brane analogue of Hawking radiation, and information loss; D-branes
as probes of black holes; and the Matrix theory approach to charged
and neutral black holes.   Some introductory material is included.

\vfill

\begin{center}
To be published as a Topical Review in {\it Classical and
Quantum Gravity}.
\vskip0.25truein
December 30, 1997.
\end{center}

\newpage
\section{Introduction}\label{intro}

Studies of black hole physics during the late 1960's and early 1970's
yielded laws of black hole mechanics \cite{jmbbcswh} that bear a
striking resemblance to the laws of thermodynamics.

\medskip
\begin{tabular}{|l|l|l|} \hline
Law &  Thermodynamics  & Black Holes \\ 
  \cline{1-3} 
Zeroth & Temperature $T$ constant  & 
Surface gravity $\kappa$ constant \\ 
 & over body in thermal equilibrium & 
over horizon of stationary black hole \\  
  \cline{1-3} 
First & $dE = T dS +$ work terms 
& $dM = \kappa dA /8\pi +$ rotation, charge terms \\ 
  \cline{1-3} 
Second & $\delta S \geq 0$ in any process &
$\delta A \geq 0$ in any process \\  
  \cline{1-3} 
Third & Impossible to achieve $T=0$ &
Impossible to achieve $\kappa =0$ \\
 & via physical processes & via physical processes  \\ \hline
\end{tabular}
\medskip

\noindent Motivated by these parallels, and the fact that the
irreducible mass \cite{dc} of a black hole is related to the area of
the horizon, Bekenstein \cite{jb} argued that the entropy of the black
hole should be directly proportional to the area of the event horizon
measured in Planck units.  A starting point was the no-hair theorem
for gravity coupled to a Maxwell field, which states that the only
information available to observers outside a black hole is a set of
conserved quantities: the ADM mass $M$, the charge $Q$, and the
angular momentum $J$.  The presence of the black hole event horizon
thus motivated Bekenstein's definition of the black hole entropy as a
measure of information about the black hole interior that is
inaccessible to outside observers.  Bekenstein then argued, on the
basis of gedankenexperiments involving infalling particles and then
merging black holes, that the entropy-area relation should be linear.

Bekenstein also proposed a generalized second law in which the total
entropy, given by the sum of the entropy of the black hole and the
common entropy of outside stuff, is nondecreasing.  This generalized
second law can be used to argue that the entropy of a self-gravitating
system of a given spatial extent and given $(M,Q,J)$ can never exceed
that of a black hole.  Alternatively phrased in terms of information
theory, this says that the maximum amount of information in a
self-gravitating system is, up to a constant of order unity, a single
bit per unit Planck area.  Although Bekenstein's arguments were made
in four dimensions, they carry over to other dimensions as well.

Hawking \cite{swh} discovered that black holes emit radiation, due to
quantum pair production in their gravitational potential gradient and
the presence of the event horizon.  The emitted radiation has a
thermal spectrum, with deviations from a perfect blackbody spectrum,
the greybody factors, determined by the frequency dependence of the
gravitational potential barriers outside the event horizon.  The
thermodynamic temperature is given in terms of the geometrical surface
gravity at the event horizon,
\begin{equation}
T_H = {\frac{\hbar\kappa}{2\pi}} \quad .
\end{equation}
An immediate consequence of this identification of the temperature is
that the proportionality constant between the entropy and the area is
fixed\footnote{We have used units in which $k_B\!=\!c\!=\!1$.  We now
set $\hbar\!=\!1$ also.}
\begin{equation}\label{bhentropy}
S_{BH} = {\frac{A}{4\hbar G_d}} \quad ,
\end{equation}
where $G_d$ is the $d$-dimensional Newton constant.  Notice that this
entropy relation is quite different from the relation for a
non-gravitating system, where the entropy scales as the volume.

Hawking's subsequent conjecture that the thermal spectrum of black
hole radiation implies information loss in quantum theory created a
puzzle which is unsolved even today.  The approximations used in
finding the entropy and temperature of black holes were semiclassical.
Physicists who found the idea of loss of unitarity unacceptable thus
looked toward better approximation schemes for a way out.  Over the
years, various attempts were made to improve the approximations, such
as including some backreaction of the black hole geometry
\cite{cgcsbgjahans} when stuff is thrown into a black hole.

The other puzzle remaining was the unknown statistical origin of the
thermodynamic black hole entropy.  A semiclassical calculation of the
partition function of Euclidean quantum gravity was performed in
\cite{gwgswh}, and yielded the Bekenstein-Hawking entropy.  However,
it was difficult to justify ignoring quantum corrections.  In
addition, the continuation from Lorentzian to Euclidean signature in
quantum gravity is not well understood.  The nature of the microscopic
degrees of freedom giving rise to the black hole entropy was therefore
still obscured from view.

The entropy and information puzzles are connected
\cite{ls9309,lsjru394,g'th,bekmg7}.  {}From study of both of them, it
gradually became clear that their resolutions may be found only in a
fully unified quantum theory of gravitation and matter.  The leading
candidate for a consistent, dynamical theory of quantum gravity is
string theory.

String theory has a massless spectrum including the graviton, and at
low energy it gives supergravities as effective theories.  Black holes
therefore appear as classical solutions of low energy string theory.
These classical spacetimes receive corrections where curvatures are
large, {\em e.g.} at the classical singularity, from essentially
stringy degrees of freedom.

The perturbative massive spectrum of string theory is an infinite
tower of states, with the energy gap between adjacent states set by
the tension of the string.  The degeneracy at a given mass level
increases rapidly with mass, and depends on other quantized conserved
quantities such as charge and angular momentum.  Black hole entropy,
which in a true quantum theory of gravity should arise from the
logarithm of a degeneracy of states, also increases with mass.  It was
therefore proposed that black holes might be identified with string
states \cite{g'th2,ls9309} and thus the entropy could be
calculated \cite{ls9309,lsjru39495,lsjr}.  This proposal turned out to
be incomplete because the logarithm of the string degeneracy did not
agree with the entropy of the black hole over the whole parameter
space unless some unknown strong-coupling physics was invoked.

The first attempt to precisely match the black hole and string state
entropies was made for extremal electric black holes in heterotic
string theory in \cite{ass}.  An obstruction which arose was that
black holes with only electric charges have zero classical horizon
area and thus zero entropy.  This problem was sidestepped via the
assumption that stringy corrections smear out the horizon area of
these black holes to order unity in string units.  The entropy
calculated at this ``stretched horizon'' \cite{lsltjru3} then scales
correctly to agree with the logarithm of the degeneracy of fundamental
string states \cite{ass}.

Nonperturbative aspects of string theory have received much attention
during approximately the last three years.  Among new discoveries were
dualities relating different string theories to one another and to an
eleven dimensional theory known as M theory (see {\em e.g.}
\cite{cmhpkt,ed95,jhstasi}), D-branes (see {\em e.g.}
\cite{sccvjjp,joetasi}), and more recently a proposal for the
nonperturbative definition of M theory via a Matrix model now known as
Matrix theory (see {\em e.g.} \cite{tommatrix,biglenny}).

These discoveries have led to a great deal of activity in the
direction of black holes.  Many new classical black hole spacetimes
have been constructed and classified.  Most spectacularly, for the
first time there has been progress in identifying the microscopic
degrees of freedom responsible for the Bekenstein-Hawking entropy
\cite{anscv}.  Recent developments in understanding the nature of the
microscopic degrees of freedom behind the Bekenstein-Hawking entropy
of black holes will be the subject of this review.

We begin in Section \ref{bran} with a discussion of the supergravity
actions in ten dimensions and the kinds of $p$-brane classical
solutions arising from them.  We mention how black holes arise upon
dimensional reduction of branes.  We then discuss the supersymmetry
bound, the different types of supersymmetric (BPS) $p$-branes, their
connection to eleven dimensional BPS M-branes, and how their
gravitational fields behave when the gravitational coupling becomes
weak.  We then briefly introduce D-branes and the supersymmetric gauge
theory actions describing their dynamics, and mention T- and
S-duality.

Section \ref{corrprin} is concerned with the correspondence principle.
We first discuss black holes with two electric NS-NS charges, and
their correspondence to excited closed strings.  We then move on to
black holes with one R-R charge, and their correspondence to open or
closed strings.  We also comment on how the NS-NS and R-R
correspondences are related to one another, on the transition between
arrays and products, and on the endpoint of Hawking evaporation of a
neutral black hole.

In section \ref{bpsnear} we discuss the very successful D-brane
computation of the microscopic entropy corresponding to BPS and
near-BPS black holes.  We concentrate mainly on $d\!=\!5$ black holes
in maximal supergravity, and discuss fractionation.  We next remark on
${\cal{N}}\!=\!2,d\!=\!4$ black holes and the microscopic entropy
computation from the point of view of M-branes.  We then turn to
D-brane analogues of near-BPS black holes in $d\!=\!5$.

Section \ref{emis} is a discussion of the D-brane analogue of Hawking
radiation.  We discuss successes and some difficulties of the
effective string model for emission and absorption, and how in one
system a discrepancy is fixed via the correspondence principle.  We
then discuss nonrenormalisation arguments for near-BPS systems, and
the information puzzle.

In section \ref{dprobmatr}, we discuss D-branes as probes, and Matrix
theory.  We discuss D-probes from both the supergravity and
supersymmetric gauge theory viewpoints.  We then give a short
introduction to Matrix theory, and move to the computation of the
Matrix theory microscopic entropy for $d\!=\!5$ BPS black holes.
Finally, we discuss recent studies of neutral black holes in Matrix
theory.

We end with a brief outlook for the future and acknowledgements in
section \ref{outack}.  A Note Added at the very end contains a few
remarks on the AdS/CFT Correspondence.

\section{Branes}\label{bran}

There are five different superstring theories in $d\!=\!10$.  We will
concentrate on Type II theories, which possess
$d\!=\!10,{\cal{N}}\!=\!2$ supersymmetry, because the associated
supergravities have the largest variety of extended objects relevant
to black hole applications.

\subsection{Supergravity Actions and $p$-branes}

The massless modes of Type-II strings separate into two sectors: NS-NS
and R-R.  The two Type-II theories are distinguished by the relative
chiralities of the two supersymmetry generators; the IIB theory is
chiral while IIA is not.  The NS-NS sector is common to both and
contains the metric $G_{\mu\nu}$, antisymmetric tensor potential
$B_{(2)}$, and the dilaton $\phi$.  The R-R sector for each theory
contains antisymmetric tensor potentials $C_{(n)}$ which are of even
degree for IIB and odd for IIA.

We give here the ten dimensional low-energy effective actions for the
bosonic fields only, in the conventions\footnote{Antisymmetrization is
done with weight one, and indices are suppressed.  The signature of
spacetime is mostly minus.} of \cite{eabcmhtmo}.  For IIA the
independent R-R potentials are $C_{(1)},C_{(3)}$ and the action is
\begin{eqnarray}\label{sIIA}
S_A &=& {\frac{1}{2\kappa_{10}^2}} \int d^{10}x \sqrt{-g} \left\{
e^{-2\phi} \left[ R + 4\left(\partial\phi\right)^2 - {\frac{3}{4}}
\left(\partial B_{(2)} \right)^2 \right] \right. \nonumber\\ & & \quad
+ \left. {\frac{1}{4}} \left(2 \partial C_{(1)}\right)^2 +
{\frac{3}{4}} \left(\partial C_{(3)} - 2 \partial B_{(2)} C_{(1)}
\right)^2 \right\} + {\frac{1}{64}} \epsilon \partial C_{(3)} \partial
C_{(3)} B_{(2)} \quad .
\end{eqnarray}
For IIB the R-R potentials are $C_{(0)},C_{(2)},C^{(+)}_{(4)}$ and the
action is\footnote{For the sake of brevity, we have suppressed
subtleties concerning the self-dual five-form field strength; for
details in resolving these, see \cite{sd5f1,sd5f2,sd5f3}.}
\begin{eqnarray}\label{sIIB}
S_{B} &=& {\frac{1}{2\kappa_{10}^2}} \int d^{10}x \sqrt{-g} \left\{
e^{-2\phi} \left[ R + 4\left(\partial\phi\right)^2 - {\frac{3}{4}}
\left(\partial B_{(2)}\right)^2 \right] \right. \nonumber\\ & & \quad
- \left. {\frac{1}{2}} \left(\partial C_{(0)}\right)^2 - {\frac{3}{4}}
\left(\partial C_{(2)} - C_{(0)} \partial B_{(2)}\right)^2 +
{\mbox{(self-dual five-form)}} \right\} \quad .
\end{eqnarray}
In these actions, we have set 
\begin{equation}
\alpha^\prime \equiv \ell_s^2 = 1 \quad ,
\end{equation} 
where $\tau_{F1}\!=\!1/(2\pi\alpha^\prime)$ is the tension of the
fundamental string.  The gravitational coupling $\kappa_{10}$ is given
in terms of the closed string coupling $g$ by
\begin{equation}\label{kappag}
2\kappa_{10}^2 =  16 \pi G_{10}= (2\pi)^7 g^2 \ell_s^8 \quad .
\end{equation}

By analogy with point-particle couplings in electromagnetism, extended
objects can couple to R-R and NS-NS antisymmetric tensor fields.
Extended objects of spatial dimension $p$ are known as $p$-branes, and
they thus have a $(p\!+\!1)$ dimensional worldvolume.  Electric
$p$-branes couple to $(p\!+\!1)$-form potentials, while magnetic
$p$-branes couple to $(7\!-\!p)$-form potentials.  Their classical
spacetimes, the black $p$-branes, have horizons and are asymptotically
flat \cite{gthans91}.  Thus in ten dimensions, we may have NS-NS
strings (NS1) and fivebranes (NS5) which couple to $B_{(2)}$, and R-R
$p$-branes (R$p$) for $p\!=\!0,2,4,6$ for IIA and $p=1,3,5$ for IIB
which couple to $C_{(n)}$.  There is also a gravitational wave W, and
the Kaluza-Klein monopole KK6 with NUT charge.  R$p$-branes for
$p=7,8,9$ are not asymptotically flat; in these cases there are too
few dimensions transverse to the $p$-brane for a Coulomb field to fall
off as a power law.  For an early review of extended objects in string
theory, see \cite{mjdrrkjxl}.

String theories in dimensions below ten are obtained by compactifying
unwanted directions on small manifolds.  Compactification of Type-II
theories on tori produces maximally supersymmetric theories, such as
${\cal{N}}=8$ supergravity in four dimensions, while compactification
on less supersymmetric manifolds produces lower dimensional theories
with fewer supersymmetries.  For compactifications on tori, the
actions in lower dimensions may be derived by applying the
Kaluza-Klein procedure to the ten dimensional action, and symmetries
of these theories may be seen \cite{jmjhs,sen9402}.  For bosonic
Lagrangians for all lower dimensional maximal supergravities via
dimensional reduction from $d=11$, see \cite{hlcp}.  A complete
classification of supergravities in diverse dimensions may be found in
\cite{ases}.

The Newton constant in dimensions lower than ten is obtained from the
ten dimensional quantity by dimensional reduction.  It is
\begin{equation}\label{Gd10}
G_d = {\frac{G_{10}}{(2\pi)^{10-d}V_{10-d}}} \quad ,
\end{equation}
where $(2\pi)^{10-d}V_{10-d}$ is the volume of the compactification
manifold.  In $d$ dimensions, the Newton constant has units of
(length)$^{d-2}$.  We define the Planck length in $d$ dimensions,
$\ell_d$, via
\begin{equation}\label{elld}
16\pi G_d \equiv 2\kappa_d^2 \equiv (2\pi)^{d-3} \ell_d^{d-2} \quad{.}
\end{equation}

Note that the relation (\ref{Gd10}) implies a neat consistency of the
Bekenstein-Hawking formula (\ref{bhentropy}) in various dimensions.
For example, we may consider a $d$-dimensional black hole also as a
$(d\!+\!p)$ dimensional $p$-brane, by uncompactifying $p$ dimensions.
Then the relation (\ref{Gd10}) ensures that the entropy is the same
whether it is calculated in the higher or lower dimension:
\begin{equation}
S_{BH} = {\frac{A_d}{4 G_d}} = {\frac{A_{d+p}}{4 G_{d+p}}} \quad .
\end{equation}

Upon dimensional reduction, tensors in both the NS-NS and R-R sectors
give rise to lower dimensional gauge fields, whose charges may be
carried by black holes.

Methods are available in string theory for generating new black hole
and $p$-brane solutions algebraically from known ones \cite{ass91}.
Solutions so constructed may be straightforwardly checked against the
differential equations of motion.  For black holes in Type-II string
theory, these methods may be applied directly by starting from the
higher dimensional generalizations of Kerr black holes \cite{rcmmjp}.
For a recent comprehensive review of new solutions in various
dimensions, methods for constructing them, $p$- and M-brane origin of
black hole geometries, classifications, intersection rules, branes at
angles, and references, see \cite{dy}.

\subsection{Supersymmetry and the String Coupling}

Of central importance in a supergravity theory is the supersymmetry
algebra, which describes how supersymmetry generators intertwine with
one another and with generators of the Poincar{\'{e}} group.  In fact,
anticommutators of the supergenerators $Q_\alpha$ give back not only
the momentum but can also produce central terms.  Schematically,
\begin{equation}
\{ Q_\alpha, Q_\beta \} \sim \left( C \Gamma^\mu \right)_{\alpha\beta}
P_\mu + \left( C \Gamma^{\mu_1\ldots\mu_p} \right)_{\alpha\beta}
Z_{[\mu_1\ldots\mu_p]} \quad ,
\end{equation}
where $C$ is the charge conjugation matrix and $\Gamma$'s are
antisymmetric combinations of gamma matrices.  {}From this we see that
objects carrying $Z$-charge are $p$-dimensional extended objects, {\em
i.e.} $p$-branes.  A recent survey of supergravities in various
dimensions and the kinds of black objects that can carry various
central charges, relevant to D-brane comparisons, may be found in
\cite{sfjmm}.  For a discussion of the importance of the superalgebra
in the M theory context see \cite{pkt9712}.
 
If we choose the rest frame, sandwiching physical states around the
above relation gives rise to a positivity bound on the mass per unit
$p$-volume of a $Z$-carrying state in that theory.  This bound may be
represented schematically by
\begin{equation}\label{bpsbd}
M \geq c |Z| \quad ,
\end{equation}
where the constant $c$ depends on the theory and its couplings.  By
inspecting the positivity bound, we see that there are special states
in this theory, as in any supersymmetric theory, which saturate the
bound.  They are known as BPS states and preserve some fraction of the
supersymmetry of the theory.  A supersymmetric nonrenormalisation
theorem protects the mass-charge equality from quantum corrections.
There is also a nonrenormalisation theorem that protects the
degeneracy of BPS states with given $Z$ from quantum corrections if
couplings are varied adiabatically \cite{doew}.

The Reissner-Nordstrom black hole solution can be embedded into {\em
e.g.} Type-II supergravity.  When so embedded, the zero-temperature
extremal Reissner-Nordstrom black hole is also supersymmetric; this
can be seen by inspecting the supersymmetry variations of the
supergravity fields.  The link between extremality and supersymmetry
extends to many known extremal black $p$-branes of superstring theory.
There are exceptions, however, such as rotating black holes in $d=4$,
and extremal but non-BPS black holes.  

Superstring theories are all related via dualities to an eleven
dimensional theory known as M theory.  The low energy limit of M
theory, $d=11$ supergravity, is related
\cite{dhis,cmhpkt,pkt9501,ed95} to $d=10$ IIA string theory via
compactification of the eleventh\footnote{We use Townsend's convention
for labeling the eleventh coordinate $x^\natural$.}  coordinate
$x^\natural$ on a circle of radius
\begin{equation}\label{Rnatural}
R_\natural = g \ell_s \quad ,
\end{equation}
where $g$ is the string coupling.  At weak coupling this is a very
small radius and the eleventh dimension is invisible; at strong
coupling it opens up and becomes large.  The length scale associated
to $d\!=\!11$ supergravity is the eleven dimensional Planck length,
which from (\ref{kappag},\ref{elld},\ref{Rnatural}) is
\begin{equation}\label{ell11}
\ell_{11} = g^{1/3}\ell_s \quad .
\end{equation}

BPS $p$-branes play a special r{\^{o}}le in duality relations: because
their mass-charge relationship is unchanged by quantum corrections, we
can follow them into strong coupling.  The BPS $p$-branes occurring in
Type-II superstring theories then have an M theory interpretation
\cite{pkt9501}; $p$-brane solutions of $d=11$ supergravity were first
found in \cite{guven}.  There are four basic BPS M-theory objects: the
gravitational wave (MW), the membrane (M2), the fivebrane (M5) and the
KK monopole (MK).  Of these, the MW and MK are purely gravitational,
while the M2-(M5-)brane carries electric (magnetic) charge associated
to the 3-form antisymmetric tensor potential of $d=11$ supergravity.

Consider the BPS M2.  In $d\!=\!11$ it may extend in the $x^\natural$
direction or not; these are termed longitudinal and transverse
M2-branes respectively.  If the M2 is longitudinal, then in
weak-coupling IIA string theory we see a one-dimensional object; if it
is transverse, we see a two-dimensional object.  In this way, M-theory
BPS objects give rise to IIA BPS objects: the M2 gives rise to the NS1
and R2 of IIA, the MW gives rise to R0 and W, the M5 to R4 and NS5,
and the MK to R6 and KK6.

Let us now consider the bound (\ref{bpsbd}) in the context of Type-II
supergravities.  The constant $c$ in the bound is different for
different BPS charge-carrying objects \cite{ed95,cmhpkt}:
\begin{equation}\label{cnsr}
c_{NS1} \sim 1 \quad , \qquad c_{Rp} \sim {\frac{1}{g}} \quad , \qquad
c_{NS5} \sim {\frac{1}{g^2}} \quad .
\end{equation}
{}From this we see that the NS1 is a fundamental object, while there
are two qualitatively different kinds of solitonic objects.

Corrections to the flat metric for a BPS black $p$-brane are of the
form \cite{gthans91} $\delta G_{\mu\nu} \sim G M / r^{7-p}$.  For the
NS1- and R$p$-branes, from (\ref{kappag},\ref{bpsbd},\ref{cnsr}) we
see that at fixed charge $|Z|$ corrections to the flat metric vanish
as we turn off the string coupling $g\rightarrow 0$ ($\ell_s\!=\!1$):
\begin{equation}\label{sugravary}
\delta G^{[NS1]}_{\mu\nu} \sim g^2 |Z| {\frac{1}{r^{7-p}}} \rightarrow
0 \quad , \qquad \delta G^{[Rp]}_{\mu\nu} \sim g^2 \frac{|Z|}{g}
{\frac{1}{r^{7-p}}} \rightarrow 0 \quad .
\end{equation}
For the solitonic NS5-brane, $c_{NS5} \sim 1/g^2$ and the metric does
not become flat for $g\rightarrow 0$; likewise for the solitonic
Kaluza-Klein monopole.

For NS1 and R$p$ branes, at weak coupling, the deviations from flat
spacetime are thus negligible.  We may then ask what describes the
dynamics of these branes at weak coupling.  The weak-coupling sister
of the BPS NS1-brane is the fundamental string F1 \cite{adjah,dghrr}.
The weak-coupling sisters of the BPS R$p$-branes are the D$p$-branes
\cite{jdrgljp,rgl,jp95}.  We now turn to the subject of D-brane
dynamics.

\subsection{D-branes}

D-branes are hypersurfaces on which open strings end; for the
definitive introduction to D-branes, see \cite{sccvjjp,joetasi}.  A
single D$p$-brane carries unit R-R charge \cite{jp95}.  Since D-branes
are BPS, two static D-branes exert zero force on one other, because
gravitational and dilatonic attraction is cancelled by electrostatic
repulsion.  So large charges can be built up by stacking many
D$p$-branes on top of one another.

The dynamics of D-branes is determined by the perturbative dynamics of
the open strings ending on them.  At low energy, only the massless
modes of the open strings are relevant; higher massive modes of the
open strings decouple and are absent from the low energy effective
theory.  The massless bosonic mode of the open string is a gauge
potential.  Now, the endpoints of open strings can carry gauge labels,
known as Chan-Paton factors, at no energy cost.  This gives rise to a
gauge theory \cite{edstrings} on the D-branes; for $Q$ D-branes the
group is $U(Q)$, and the gauge fields are in the adjoint
representation.

At this point we mention dualities, which are important in the study
of D-branes and other objects in string theory.  The basic idea of a
duality is that theory ${\cal{A}}$, with its fields and coupling
constants, is related to theory ${\cal{B}}$, with its own fields and
coupling constants, by the duality transformation.  Sometimes theory
${\cal{A}}$ and theory ${\cal{B}}$ are the same, and the duality
transformation is then a symmetry of that theory.  We now mention two
duality transformations which will be of interest to us, S-duality and
T-duality.  S-duality is a symmetry of type IIB and acts in $d=10$ as
\begin{equation}\label{sduality}
S: \qquad {\tilde{g}} = 1/g \quad , \qquad {\tilde{g}}^{1/4}
{\tilde{\ell}}_s = {g}^{1/4} \ell_s \quad.
\end{equation}
S-duality switches R1 with NS1, R5 with NS5, and leaves W invariant
while R3 is self-dual.  T-duality is a symmetry of all closed string
theories and acts on a compact direction, say $x^9$, with radius
$R_9$.
\begin{equation}\label{tduality}
T: \qquad {\frac{{\tilde{R}}_9}{{\tilde{\ell}}_s}} =
{\frac{\ell_s}{R_9}} \quad , \qquad
{\frac{{\tilde{g}}}{\sqrt{{\tilde{R}}_9/{\tilde{\ell}}_s}}} =
{\frac{g}{\sqrt{R_9/\ell_s}}} \quad , \qquad {\tilde{\ell}}_s = \ell_s
\quad .
\end{equation}
Acting on fundamental strings, T-duality switches winding and momentum
modes.  On D$p$-branes, T-duality changes the dimension of the
worldvolume by $\pm\!1$, depending on whether or not it acts in a
direction perpendicular to the worldvolume.  This is in fact one way
to see why having an open string sector in a closed string theory
requires D-branes \cite{sccvjjp,joetasi}.  Application of T-duality to
systems with nontrivial gravitational fields is more subtle; {\em
e.g.} for the NS5 we have to smear out the dependence on a transverse
coordinate if we want to apply T-duality in that direction.  If we
then T-dualize, we get the KK6.  In a similar way, if we turn on the
R$p$-brane gravitational fields, we have to smear on a transverse
direction before T-dualizing it.

The low-energy effective action for a test D-brane in a supergravity
background is in string units \cite{kappa1,kappa2,eabpkt,macpjhs}
\begin{equation}\label{sdbiwz}
I =  - \tau_{Dp} \int d^{p+1}\sigma \  e^{-\phi}
\sqrt{-\det\left(G_{\alpha\beta} + 
\left[2\pi F_{\alpha\beta}-B_{\alpha\beta}\right]\right)} 
 - \tau_{Dp} \int e^{2\pi F-B} \wedge \bigoplus_n C_{(n)} \quad ,
\end{equation}
where the D-brane tension is \cite{sccvjjp,joetasi}
\begin{equation}\label{dbtension}
\tau_{Dp} = {\frac{\sqrt{\pi}}{\kappa_{10}(2\pi\ell_s)^{p-3}}}  
       = {\frac{1}{(2\pi)^p g \ell_s^{p+1}}} \quad .
\end{equation}
The action is gauge-invariant, and it is supercovariant in spacetime:
all fields are superspace extensions of the ordinary bosonic fields
and they are pulled back to the bosonic worldvolume in a spacetime
supersymmetric fashion.  Suppressing fermions, an example is
$G_{\alpha\beta} = G_{\mu\nu} \partial_\alpha X^\mu \partial_\beta
X^\nu$.  For the lowest components of the superfields, $C_{(n)}$ are
the R-R potentials, $B_{(2)}$ is the NS-NS potential, $F_{(2)}$ is the
worldvolume gauge field, and $G_{\mu\nu},\phi$ are the metric and
dilaton.  Kappa-symmetry of the above action
\cite{kappa1,kappa2,eabpkt,macpjhs} puts the supergravity background
on-shell.  

At weak coupling, the bulk (supergravity) fields decouple and we are
left with a gauge theory on the brane.  At weak fields, the
Born-Infeld (first) term in the action is then to lowest order $U(1)$
supersymmetric gauge theory.  For $Q$ D-branes, the weak-field theory
is $U(Q)$ supersymmetric Yang-Mills (SYM).  The field content and
Lagrangian are determined by dimensional reduction from the $d=10$
$A_\mu$, and so on the brane we have a gauge potential $A_\alpha$ and
$(9\!-\!p)$ scalar fields $X^i\equiv -A_i$.  The bosonic Lagrangian is
then in string units
\begin{equation}\label{axax}
S_{\mbox{\footnotesize{bos}}} \sim \int {\frac{1}{g}} \mbox{tr}
\left\{ - {\frac{1}{4}}F^{\alpha\beta}F_{\alpha\beta} + {\frac{1}{2}}
D^\alpha X^i D_\alpha X^i + {\frac{1}{4}}\left[X^i,X^j\right]^2
\right\} \quad{.}
\end{equation}
When the branes are pulled apart, the gauge group is broken to
$U(1)^Q$ and the $Q$ diagonal entries in the $Q\times Q$ matrices
$X^i$ are then the positions of the individual D-branes in the $i$th
direction.  Higher order Born-Infeld corrections in the gauge theory
effective action may in principle be found from string scattering
amplitude calculations; see \cite{aatdbi} and \cite{dbmjp} for the
nonabelian generalisation of the Born-Infeld term.

In addition, the Wess-Zumino (second) term allows smaller branes to
live inside larger ones \cite{mrd}, and branes to end on branes
\cite{ans9512,pkt9512}; for a nice review of the latter from the M
theory perspective, see \cite{pkt9609}.  An example of a configuration
which is supersymmetric and has zero binding energy is
\cite{sccvjjp,joetasi} a D$(p\!-\!4)$-brane living inside a
D$p$-brane.  In this case the smaller brane creates worldvolume gauge
fields in the SYM theory on the bigger brane which are analogous to
(zero-size) instantons \cite{mrd}.

D-branes are BPS and may be combined by using relatively simple
recipes to make intersecting configurations, which break more
supersymmetry than plain branes.  The overall fraction of
supersymmetry preserved may be worked out for boundstates with zero
binding energy by imposing each brane's spinor projection condition in
turn, and checking consistency conditions on spinors in a pairwise
fashion.  For branes at angles and configurations dual to them, the
supersymmetry analysis is a little less straightforward \cite{bdl}.

Now, since the dynamics of D-brane configurations at low energy and
weak coupling is given in terms of supersymmetric gauge theories, it
is possible in some cases to compute the degeneracy of states for a
configuration with given quantum numbers.  Since D-branes are BPS
states, supersymmetry also protects their degeneracy for fixed quantum
numbers from quantum corrections as we move adiabatically from weak to
strong coupling.  This implies \cite{anscv} that the entropy of a
bunch of D-branes should be the same as the entropy of the analogous
supersymmetric black $p$-brane configuration.  In fact, this idea has
a generalisation to non-BPS systems as well.

\section{The Correspondence Principle}\label{corrprin}

Classical supergravity black hole solutions have length scales
associated to them, connected to the parameters of the solution, which
determine the length scales over which gravitational fields vary
noticeably.  These length scales typically depend on $g$, due to
(\ref{cnsr}).

As we will see in detail in later sections, R-R $p$-brane supergravity
spacetimes are valid in the regime where these gravitational radii are
large in string units, while the D-brane picture is valid in the
opposite limit.  This complementarity of pictures also holds for the
NS1-brane, whose weak-coupling sister is the F1, and thus for many
black objects in string theory.  We may ask, therefore, if there is
some regime where the D-brane/string and supergravity descriptions
turn into one another.

An early idea related to the correspondence principle was discussed
for four dimensional Schwarzschild black holes in \cite{ls9309}.  The
Black Hole Correspondence Principle of \cite{gthjp} states that, upon
turning up the string coupling $g$, a D-brane/string state will turn
into a black hole when the curvature at the horizon of the
corresponding black hole is of order the string scale.  This is called
the correspondence point.  The striking observation of \cite{gthjp}
was that if we demand that the masses and other conserved quantum
numbers of the two different configurations match at the
correspondence point, then the entropies also match.  Note that this
correspondence generically occurs only at a point in parameter space,
and so it resolves the difficulty previously found in early attempts
to match string state and black hole entropy.  In this section we
review the correspondence principle of \cite{gthjp}; see also the
review \cite{gth9704}.

\subsection{Two Electric NS-NS Charges}

Generic charged black holes in string theory have a position-dependent
dilaton field.  The string feels a different metric $g_{\mu\nu}$ to
the Einstein metric $g_{\mu\nu}^E$; they are related by
\begin{equation}
g^E_{\mu\nu} = e^{-4\phi/(d-2)} g_{\mu\nu} \quad .
\end{equation}
In applying the correspondence principle, we must specify which frame
we use in calculating the curvature.  The correct frame to use is the
natural one, the string frame.

The metric for a $d$-dimensional electrically charged NS-NS black hole
is \cite{awp} in string frame\footnote{We have changed notation
slightly from that work: $\alpha,\beta=\alpha_p \pm \alpha_w$.  We
have also picked the charge vector to lie along the $9$ direction.}
\begin{equation}\label{nsbh}
ds_d^2 = {\frac{\left[1-k(r)\right]}{f_p(r) f_w(r)}} dt^2 -
{\frac{dr^2}{\left[1-k(r)\right]}} - r^2 d\Omega_{d-2}^2 \quad ,
\end{equation}
where
\begin{eqnarray}
k(r)   &=& \left({\frac{r_H}{r}}\right)^{d-3}  \quad ,\\
f_p(r) &=& 1 + k(r) \sinh^2\!\alpha_p  \quad , \\
f_w(r) &=& 1 + k(r) \sinh^2\!\alpha_w  \quad .
\end{eqnarray}
In these equations, $r_H$ is the horizon radius.  The $\alpha$ are
rapidities corresponding to the algebraic boost transformations used
to generate the solutions; without loss of generality we can take them
to be positive.  The dilaton, and internal modulus field along the
string direction, are given by \cite{awp}
\begin{eqnarray}
e^{-4\phi} &=& f_p(r) f_w(r) \quad , \\
G_{99}     &=& f_p(r)/f_w(r) \quad . \label{modns}
\end{eqnarray}
The ADM mass and gauge charges are \cite{awp}
\begin{eqnarray}
M &=&           {\frac{(d-3)\omega_{d-2}r_H^{d-3}}{16\pi G_d}} \left[
{\frac{1}{(d-3)}} + {\frac{1}{2}}\left( \cosh\!2\alpha_p + \cosh\!
2\alpha_w \right) \right] \label{mns} \quad , \\
q_{p,w} &=& {\frac{(d-3)\omega_{d-2}r_H^{d-3}}{16\pi G_d}} \left[
{\frac{1}{2}}\sinh\!2\alpha_{p,w} \right] \quad , \label{qns}
\end{eqnarray}
where $\omega_n={\mbox{Vol}}(S^n)=
2\sqrt{\pi}^{n+1}/\Gamma({\frac{n\!+\!1}{2}})$.  Classically the
rapidities $\alpha$ are continuous parameters, but in the quantum
theory the following quantities are integer normalised:
\begin{equation}
Q_p = q_p R_9 \quad , \qquad Q_w = {\frac{q_w}{R_9}} \quad .
\end{equation}
For neutral black holes $\alpha_{p,w}=0$.  The entropy and Hawking
temperature are \cite{awp}
\begin{eqnarray}\label{sbhns}
S_{BH} &=& {\frac{\omega_{d-2} r_H^{d-2}}{4G_d}}
\cosh\!\alpha_p\cosh\!\alpha_w \quad , \\
T_H &=& {\frac{(d-3)}{4\pi r_H \cosh\!\alpha_p\cosh\!\alpha_w}} \quad .
\end{eqnarray}

At correspondence, the black hole becomes very small and the curvature
at the horizon is large.  Then stringy corrections will kick in and
modify the metric (\ref{nsbh}); for example, the function
$\left[1-k(r)\right]$ will get smeared out to order unity.  This is
the stretched horizon phenomenon \cite{ass}.  We can also see from the
metric that the charges, via the $\alpha$'s, give rise to a redshift
of the energy, of order $\left(\cosh\!\alpha_p\cosh\!\alpha_w\right)$.
Then from (\ref{modns}) the $x^9$ direction gets contracted at the
horizon by a factor $\left(\cosh\!\alpha_w/\!\cosh\!\alpha_p\right)$
from its value at infinity.  If the charges are zero, these redshift
effects go away.

Let us now determine the correspondence point.  {}From (\ref{nsbh})
the curvature at the horizon in ten dimensional string frame is
${\cal{O}}(r_H^{-2})$, and so the correspondence point occurs at
\cite{gthjp}
\begin{equation}\label{nscorrpt}
r_H \sim \ell_s \quad .
\end{equation}

The black hole's weak-coupling sister is an excited state of a
fundamental string wrapped around $x^9$.  The string has the same
conserved quantum numbers as the black hole, at correspondence.  For
the left- and right-moving momenta of the string excitations,
\cite{mbgjhsew}
\begin{equation}
p_{L,R} = {\frac{Q_p}{R_9}} \mp {\frac{Q_w R_9}{\ell_s^2}} \quad ,
\end{equation}
while for weak coupling the entropy scales as
\begin{equation}\label{sss}
S_{SS} \sim \sqrt{N_R} + \sqrt{N_L}  \quad ,
\end{equation}
where 
\begin{equation}\label{nnn}
N_{R,L} \sim \ell_s^2 \left( M^2 - p_{R,L}^2 \right)
\quad .
\end{equation}
In determining $N_{L,R}$ we need to be careful to take account of any
redshift effects.  Redshifts will be pronounced, {\em i.e.} not
numbers of order unity, when the rapidities $\alpha$ are large.  As we
can see from (\ref{mns},\ref{qns}), large-$\alpha$ corresponds to a
near-BPS configuration.  To see how this affects determination of
$N_{L,R}$ it is simplest \cite{gthjp} to turn off $\alpha_p$; then the
energy is redshifted by the same factor as $R_9$ is Lorentz contracted
and $N_L\!=\!N_R\!\equiv\!N$.  The energy above extremality is small,
and is of order
\begin{equation}
\Delta E \sim M - {\frac{Q_w R_9}{\ell_s^2}} \sim {\frac{N}{Q_w R_9}}
\quad .
\end{equation}
{}From this we see that the effect of the redshift/contraction cancels
in determination of $N$.  Turning $\alpha_p$ back on does not affect
the determination of $N_{L,R}$ either \cite{gthjp}.

Now we are ready to find the entropy of the string state.  We use $r_H
\sim \ell_s$, together with the charge and mass matching relations and
(\ref{sss},\ref{nnn}), to find the entropy at correspondence
\cite{gthjp}:
\begin{equation}
S_{SS} \sim 
{\frac{\ell_s^{d-2}}{G_d}}\cosh\!\alpha_p\cosh\!\alpha_w \quad ,
\end{equation}
up to numbers of order unity, {\em i.e.} to the accuracy of the
correspondence principle.  This agrees with $S_{BH}$, as advertised.
Note that since the asymptotic string coupling $g$ is taken to be
weak, this entropy is large.

Let us figure out the value of the string coupling at correspondence.
We have that the internal compactification volume is small; to the
accuracy of matching scalings we can ignore its precise numerical
value.  Then, using $G_d \sim g^2 \ell_s^{d-2}$, and the fact that the
local string coupling on the horizon is $g e^{\phi(r_H)}$, we find
\begin{equation}
\left. (e^{\phi_c} g_c) \right|_{r_H} \sim {\frac{1}{N^{1/4}}}
\quad{,}
\end{equation}
which is a weak coupling.  This justifies our assumption that the
entropies of left- and right-movers is additive, as in (\ref{sss}).
We can calculate the local temperature at the correspondence point;
taking account of the redshift it is $\sim\!1/\ell_s$.

\subsection{One R-R Charge}

Application of the correspondence principle to systems with one R-R
charge shows new features so we now discuss this case.  Earlier
attempts to understand the entropy of near-extremal R-R $p$-branes may
be found in \cite{ssgirkawp,irkaat9604}; see also \cite{odin}.

The metric for a black R$p$-brane with worldvolume spatial coordinates
${\vec{y}}$ is \cite{gthans91} in string frame
\begin{equation}\label{nerrmetric}
ds_{10}^2 = f_p(r)^{-1/2}\left[\left(1-k(r)\right)dt^2 -
d{\vec{y}}^2\right] - f_p(r)^{1/2}\left[\left(1-k(r)\right)^{-1}dr^2 +
r^2 d\Omega_{8-p}^2\right] \quad ,
\end{equation}
where
\begin{equation}
k(r) = \left({\frac{r_H}{r}}\right)^{7-p} \quad , \qquad 
f_p(r) = 1 + k(r)\sinh^2\!\beta \quad .
\end{equation}
To obtain a black hole in $(10-p)$ dimensions we roll up the $p$-brane
on a manifold of volume $(2\pi)^{p}V_p$.  The resulting black hole is
known as a parallel $p$-brane black hole; there is a R-R gauge field
and the dilaton is
\begin{equation}\label{dilrr}
e^{\phi} = f_p^{(3-p)/4} \quad .
\end{equation}
The mass, R-R charge, entropy and Hawking temperature are given
by\footnote{In comparing to other references, such as
\cite{rgcrkspkny}, the reader may wish to note the following
conversions: $d=(7-p)$; $\alpha=(3-p)/2$;
$r_-^{7-p}=r_H^{7-p}\sinh^2\!\beta$;
$r_+^{7-p}=r_H^{7-p}\cosh^2\!\beta$; we have also performed the
coordinate transformation $r^{7-p} := r^{7-p} - r_-^{7-p}$.}
\begin{eqnarray}
M &=& {\frac{(7-p)\omega_{8-p}r_H^{7-p}}{16\pi G_{10-p}}} \left[
{\frac{1}{2}} + {\frac{1}{7-p}} + {\frac{1}{2}}\cosh\!2\beta \right]
\quad , \label{mrr} \\ Q &=&
{\frac{(7-p)\omega_{8-p}r_H^{7-p}}{(2\pi)^{7-p}g}} \left[
{\frac{1}{2}}\sinh\!2\beta \right] \quad , \label{qrr} \\ S_{BH} &=&
{\frac{\omega_{8-p}}{4 G_{10-p}}} r_H^{8-p} \cosh\!\beta \quad ,
\label{sbhrr} \\ T_H &=& {\frac{(7-p)}{4\pi r_H \cosh\!\beta}}
\label{thawrr} \quad .
\end{eqnarray}
Note that with these conventions the BPS limit satisfies the relation
$M/V_p = Q/g$, with $Q$ integer.

We now determine the correspondence point.  The maximum curvature at
the horizon comes from the angular part of the metric transverse to
the $p$-brane worldvolume \cite{gthjp}; setting it to be of order
unity gives
\begin{equation}\label{rcorrpt}
r_H \sim {\frac{\ell_s}{\sqrt{\cosh\!\beta}}} \quad .
\end{equation}
Here we see a new wrinkle: the correspondence size of the horizon
depends on the boost $\beta$, not just the string length $\ell_s$.
We will come back to this point later.

As for the NS-NS situation we can find the local coupling on the
horizon at correspondence; from (\ref{dilrr},\ref{kappag},\ref{qrr})
it is \cite{gthjp}
\begin{equation}\label{glrr}
\left. (e^{\phi_c} g_c) \right|_{r_H} \sim {\frac{1}{Q}}  \quad ,
\end{equation}
which is small for macroscopic charge.  For $p\!<\!3$, there is a
subtlety, that the curvature increases away from the horizon to a
maximum before it then decreases to zero at infinity \cite{gthjp}.
Now, super-stringy curvatures lead to corrections to the classical
geometry.  It is therefore interesting that taking the horizon
curvature to be the relevant one for correspondence is the right
prescription.  For $p\!\geq\!3$, there is no such subtlety; we could
imagine resolving the $p\!<\!3$ difficulty by T-duality from this
case.  We have also been informed that the correspondence principle
works for rotating black holes \cite{rcmgth}; there is a similar kind
of subtlety encountered there.

For the sister state, the first thing we need is to determine the
nature of the degrees of freedom carrying the energy.  We know that
the BPS R$p$-brane's sister is the D$p$-brane, but this only carries
the BPS energy.  We need other degrees of freedom to carry the
energy above the BPS energy,
\begin{equation}
\Delta E \equiv E - {\frac{Q V}{g \ell_s^{p+1}}} \quad .  
\end{equation}
The BPS R$p$-brane has zero classical entropy, so the other degrees of
freedom carrying $\Delta E$ must also carry the entropy for a
nonextremal $p$-brane.  There are two options \cite{gthjp}: long
(closed) strings in the plane of and near the D$p$-brane, and massless
open strings running along the D$p$-brane.

For the former, the energy $\Delta E$ is proportional to the length,
and so is the entropy, and dimensional analysis gives
\begin{equation}\label{closedstate}
S_{closed} \sim \ell_s \Delta E \quad .
\end{equation}
For the latter, the massless open string degrees of freedom are those
of a $U(Q)$ gauge theory on the brane.  There are worldvolume gauge
fields and transverse scalars, all in the adjoint representation, and
so there are $\sim\!Q^2$ degrees of freedom.  A first approximation to
this system is a free gas, and if we assign the gas a temperature $T$
we can write the excess energy and entropy as
\begin{equation}\label{openstate}
\Delta E_{open} \sim Q^2 V_p T^{p+1} \quad , \qquad
S_{open} \sim Q^2 V_p T^p \quad .
\end{equation}
There are a couple of subtleties to attend to \cite{gthjp}.  The first
is that there is a redshift.  Since all worldvolume metric components
involve $f_p^{-1/2}$, the redshift factor is $\sqrt{\cosh\!\beta}$,
but all worldvolume lengths scale inversely to the excess energy and
temperature, so the entropy in (\ref{openstate}) is unaffected.  The
closed string is in the plane of the $p$-brane, and so its entropy
(\ref{closedstate}) is also unaffected by the redshift.  This redshift
also implies via (\ref{thawrr}) that the local horizon temperature at
correspondence is of order unity in string units.

The second subtlety is that we have ignored interactions, given by
(\ref{axax}).  In perturbation theory, interaction vertices on the
brane pick up a factor of $Q$ due to Chan-Paton factors, and a factor
of the local string coupling.  Dimensional analysis then gives the
perturbation expansion parameter \cite{gthjp}: $(g e^\phi) Q T^{p-3}$.
{}From (\ref{glrr}), and the local temperature, this is of order unity.
Thus to the accuracy of the correspondence principle we can neglect
interactions.  However, if we wanted a more precise calculation we
would need to understand these interactions properly.  This estimation
of the importance of interactions illuminates a puzzle for the
threebrane found in \cite{ssgirkawp}; there, interactions were
neglected, and thus the energy was underestimated and the entropy
overestimated.

These two different kinds of degrees of freedom, the closed versus
open strings, dominate in different regimes; for large $\beta$ the
open string gas dominates and {\em vice versa}.  Matching the black
hole charge and mass to those of the closed or open string
configurations, at the correspondence point, yields via
(\ref{closedstate},\ref{openstate})
\begin{eqnarray}
S_{open}   &\sim& {\frac{V}{g^2\ell_s^p}} (\tanh\!\beta)^{2/(p+1)}
(\cosh\!\beta)^{(p-6)/2} \quad , \\
S_{closed} &\sim& {\frac{V}{g^2\ell_s^p}} (\cosh\!\beta)^{(p-7)/2} 
\quad .
\end{eqnarray}
On the other hand, the black hole entropy (\ref{sbhrr}) scales at
correspondence as
\begin{equation}
S_{BH} \sim {\frac{V}{g^2\ell_s^p}} (\cosh\!\beta)^{(p-6)/2} \quad ,
\end{equation}
which agrees qualitatively with whichever of $S_{closed,open}$ is
larger \cite{gthjp}, depending on $\beta$.

\subsection{Connections, and the Endpoint}

The correspondence between a near-BPS R$p$-brane black hole and the
open-string gas can be seen to be just the correspondence between a
neutral black hole and an excited closed string, viewed in a highly
boosted frame \cite{srdsdmskrpr}.  In showing this, use is made of the
$d\!=\!10$ IIA relation with $d\!=\!11$ M theory.  R-R black holes in
$d\!=\!(10\!-\!p)$ dimensions come from R$p$-branes wrapped around a
$p$-torus.  They are therefore related by T-duality to a (smeared)
R0-brane in $d\!=\!10$.  However, a R0-brane is obtained from a
$d\!=\!10$ Schwarzschild black hole by uncompactifying the latter to a
$d\!=\!11$ black string stretched along $x^{\natural}$, boosting in
$x^{0,\natural}$ by rapidity parameter $\beta$, and compactifying down
to $d\!=\!10$ again.  Then we convert back to the R$p$-brane black
hole by T-duality.

During the boost process, $R_\natural$ gets Lorentz contracted by the
factor $\cosh\!\beta$, while the $d\!=\!11$ Planck scale $\ell_{11}$
and the horizon parameter $r_H$ are unchanged \cite{srdsdmskrpr}.
Then from (\ref{Rnatural},\ref{ell11}) we find that the new string
length is
\begin{equation}
{\overline{\ell}}_{s} =
\left({\frac{\ell_{11}^3}{{\overline{R}}_\natural}}\right)^{1/2}=
\ell_s\sqrt{\cosh\!\beta} \quad .
\end{equation}
Using this new string length in the R-R correspondence point
(\ref{rcorrpt}), we find that it is precisely the NS-NS correspondence
point (\ref{nscorrpt}) expressed in terms of the old string length:
\begin{equation}\label{chargeneutral}
{\overline{r}}_H^{\mbox{\footnotesize{(R-R)}}} \sim
{\frac{{\overline{\ell}}_s}{\sqrt{\cosh\!\beta}}}
\quad\leftrightarrow\quad r_H^{\mbox{\footnotesize{(NS-NS)}}}\sim\ell_s
\quad .
\end{equation}
Other aspects of this mapping are discussed in \cite{srdsdmskrpr}.

Black holes in $d$ dimensions (BH$_d$) can in fact arise in two
different ways from $d\!+\!1$ dimensions.  The first is via a
$d\!+\!1$ dimensional black string, which is the direct product of the
black hole and the extra dimension $z$ of radius $R_z$.  The second is
to take an array of $d\!+\!1$ dimensional black holes spaced along $z$
by $2\pi R_z$; for a static spacetime the array must be an infinite
one \cite{rcm}.  Dimensionally reducing gives approximately BH$_d$.
If the radius of the BH$_{d+1}$ horizon satisfies $r_H/R_z \ll 1$ then
the array is the correct solution; otherwise the array becomes
unstable \cite{rgrlf} and turns into the product solution.  The reason
for this is that, if BH$_d \times S^1_z$ and BH$_{d+1}$ have the same
mass, the array has greater entropy than the product if $r_H/R_z\ll 1$
and {\em vice versa}.  It was shown in \cite{gthjp} that the
properties of D-branes and strings mesh nicely with the correspondence
principle in describing this array-product transition.

The conservative way to view the correspondence principle is to apply
it in the direction of increasing coupling $g$: a boundstate at weak
coupling of strings and D-branes turns into a black hole as we turn up
the string coupling.  For neutral black holes, the nature of the
approach in the conservative direction has been studied in
\cite{gthjp2}; the physics depends on $d$.  See also \cite{rrkpol}.

Viewed in the other direction, however, the correspondence principle
is in a sense a radical proposal.  It says that when black holes
become very small, of order the string scale, they turn into string
states.  In this way, the correspondence principle provides an answer
to the old question regarding the nature of the endpoint of Hawking
radiation of a macroscopic Schwarzschild black hole.  Namely, that the
endpoint is a highly excited string.  An interesting fact is that this
correspondence point does not occur when the black hole is of order
the Planck mass.  Taking the compactification volume to be of order
the string scale, we find the correspondence mass to be
\begin{equation}
m_{c} \sim {\frac{\ell_s^{d-3}}{G_d}} \sim {\frac{1}{\ell_sg^2}} \sim
{\frac{1}{\ell_d}} {\frac{1}{g^{2(d-3)/(d-2)}}} \quad .
\end{equation}
For weak string coupling, this is larger than the Planck mass
$1/\ell_d$; in four dimensions, the correspondence mass is larger by a
factor of $1/g$.

In this section we have not discussed the case where black holes have
more than one R-R charge on them, and possibly NS-NS charges as well.
The only really new situations occur when two or more R-R charges are
large.  In these cases, it is found that the matching scale drops out
so that exact comparisons can be done for these near-BPS (and BPS)
situations.  These precise comparisons will be the subject of the next
two sections.

\section{BPS and Near-BPS Entropy}\label{bpsnear}

As mentioned in section \ref{intro}, the earliest attempt \cite{ass}
to make precise matchings between black hole entropy and calculable
stringy degeneracies faced the obstacle that the area of the relevant
classical NS-NS charged black hole vanishes.  This problem actually
extends further: stringy BPS black holes with nonzero horizon area are
not possible for six and higher spacetime dimensions in toroidal
compactification \cite{irkaat9604} even with R-R charges included.
Roughly speaking, the physical reason for this is that there are not
enough independent charges on higher dimensional black holes for the
dilaton and moduli to have finite nonzero values at the horizon, and
this leads via the classical equations of motion to zero horizon area.

Therefore, the most interesting places to look for nonzero classical
BPS black hole entropies are in five and four dimensions.  Note in
this regard that some authors had previously argued on the basis of
semiclassical topological considerations that BPS black holes have
zero entropy.  However, a good explanation of why this semiclassical
result is unreliable in string theory may be found in \cite{gth}.
Some initial attempts to calculate the microscopic entropy of
finite-area extremal black holes from a dual perspective were made in
\cite{flfw}.

Agreement of string/D-brane and black hole entropies was first
accomplished in a seminal paper by Strominger and Vafa \cite{anscv} in
January of 1996.  There, new D-brane technology was used to count the
degeneracy of states of a D-brane system with the same quantum numbers
as a classical five-dimensional R-R charged BPS black hole.
Spectacular, exact agreement was found with the Bekenstein-Hawking
entropy.  This work was quickly followed by calculations addressing
near-extremal black holes \cite{cgcjmm,gthans9602}, rotation
\cite{jcbrcmawpcv,blmpsv}, and four dimensions
\cite{cvjrrkrcm,jmmans9603,gthdaljmm}.  Also discovered were
multiparameter agreements between semiclassical Hawking radiation and
greybody factors and analogous scattering processes directly
calculable in string theory.  Agreements between the entropy and
particle emission of various four and five dimensional black holes and
their corresponding D-brane configurations have also been reviewed in
\cite{jmmthesis,dy}, and so we refer the reader there for additional
details.

\subsection{BPS Black Holes in $d=5$}\label{bpsd5}

In five dimensional maximal supergravity, there are three independent
orthogonal charges $Q$ that a black hole can carry.  A five
dimensional black hole carrying this set of charges can be chosen via
a duality transformation to originate from a IIB $d=10$ configuration
which is the intersection of three BPS configurations: a R1-brane of
charge $Q_1$, a R5-brane of charge $Q_5$, and a gravitational wave
with momentum $Q_p$ along the intersection string.  We represent
this configuration by the matrix
\begin{equation}
{\begin{array}{l}
Q_1 \ R1 \\ Q_5\ R5 \\ Q_p \ W
 \end{array}}
\left[ 
{\begin{array}{llllllllll}
 0 & \cdot & \cdot & \cdot & \cdot & 5 & \cdot & \cdot & \cdot & \cdot 
 \\
 0 & \cdot & \cdot & \cdot & \cdot & 5 & 6 & 7 & 8 & 9 \\
 0 & \cdot & \cdot & \cdot & \cdot & 5 & \cdot & \cdot & \cdot & \cdot  
 \end{array}}
\right] \quad .
\end{equation}
Here, each line represents a constituent, and the non-dot numbers
correspond to the coordinates which involve that constituent.  For
example, the D1 is stretched along $x^5$.

This is a supersymmetric state, preserving one of eight
supersymmetries, for left-moving momentum.  The supersymmetry of this
intersecting brane configuration can be seen \cite{inter1,inter2} from
the constituent brane conditions:
\begin{equation}\begin{array}{rll}
\epsilon_L &= (\Gamma_0\Gamma_5\Gamma_6\ldots\Gamma_9) 
              \epsilon_R \quad & (R5) \quad , \\
\epsilon_L &= (\Gamma_0\Gamma_5) \epsilon_R \quad & (R1) \quad , \\
\epsilon_L &= (\Gamma_0\Gamma_5) \epsilon_L \quad ; \qquad
\epsilon_R =  (\Gamma_0\Gamma_5) \epsilon_R  
              \quad & (W) \quad ,
\end{array}
\end{equation}
where the two Majorana-Weyl spinors satisfy $\epsilon_L =
\Gamma_{\natural} \epsilon_L$ and $\epsilon_R = \eta\Gamma_\natural
\epsilon_R$ with $\eta=+1$ for IIB.  We have taken the three charges
to be positive.

The five dimensional black hole is obtained by rolling up the string
on a circle of radius $R_5$, and rolling up the fivebrane on the
direct product of the circle and a four-torus $T^4$ with volume
$(2\pi)^4 V$, $V=R_6R_7R_8R_9$.  The resulting metric is in five
dimensional Einstein frame
\begin{equation}
ds_5^2 = \left(H_1 H_5 H_p\right)^{-2/3} dt^2 - \left(H_1 H_5
H_p\right)^{1/3} \left[ dr^2 + r^2 d\Omega_3^2 \right] \quad ,
\end{equation}
where $H_i = 1 + r_i^2/r^2$ are harmonic functions and 
\begin{equation}
r_i^2 = \left. {\frac{16\pi G_d M_i}{(d-3) \omega_{d-2}}}
\right|_{d=5}\quad ,
\end{equation}
where $M_i$ is the mass of the $i$-th constituent.  We have not
written out the expressions for the gauge, dilaton, or internal moduli
fields.  {}From the gauge fields come the relations
\begin{equation}\label{r15p}
r_1^2 = {\frac{g Q_1\ell_s^6}{V}} \quad , \qquad 
r_5^2 = g Q_5 \ell_s^2 \quad , \qquad
r_p^2 = {\frac{g^2 Q_p \ell_s^8}{R_5^2 V}} \quad .
\end{equation}
These parameters are the gravitational radii associated to the three
charges on the black hole.  For the three gravitational radii to be
comparable for weak bulk coupling $g$, we need
$Q_1\!\sim\!Q_5\!\ll\!Q_p$, when $R_5,V^{1/4}\!\sim\!\ell_s$.  If all
of these scales are smaller than the string scale, the supergravity
fields are negligible.  In this regime we do not really have a black
hole; to get one, we need to grow some gravitational radii above the
string scale.

In order for the classical supergravity solution to be valid, we need
that closed string loop corrections, which are proportional to $g^2$,
be small.  We can arrange this at fixed $R_5,V$ while retaining
nontrivial gravitational fields by taking the limit \cite{jmmans9609}
\begin{equation}\label{loopcorr}
g \rightarrow 0 \quad , \qquad
Q_1, Q_5, Q_p \rightarrow \infty \quad , \qquad
r^2_{1,5,p}  \quad {\mbox{fixed}} \quad .
\end{equation}
We may also ignore stringy $\alpha^\prime$ corrections to the geometry
as long as the gravitational radii are large in string units,
\begin{equation}\label{alphacorr}
r_{1,5,p} \gg \ell_s \quad .
\end{equation}
Internal dimensions will be inaccessible to outsiders as long as
$R_5,V$ are small.  In the supergravity regime, the charges and mass
on the black hole are macroscopic in string units.

The Bekenstein-Hawking entropy associated to the black hole geometry
is given by
\begin{equation}\label{5dbhentropy}
S_{BH} = {\frac{A}{4 G_5}} = 2\pi\sqrt{Q_1 Q_5 Q_p} \quad .
\end{equation}
An important feature of this entropy is that it is independent of the
string coupling and of the moduli of the internal compactification
manifold.  This is in fact a general feature of black hole entropy;
for a recent discussion see {\em e.g.}  \cite{modent}.  In addition, the
Hawking temperature is zero, as expected for an extremal black hole.

We now turn to the weak-coupling sister D-brane configuration in order
to compute the D-brane entropy.  The first two charges are carried by
$Q_5$ D5-branes with $Q_1$ D1-branes inside.  For the third charge, we
need massless open strings with endpoints on various D1,5-branes to
carry the momentum $P\!=\!Q_p/R_5$.  The fermionic massless open
strings can also each carry $\hbar/2$ of angular momentum.  Other
configurations have also been studied which are U-dual to this one
\cite{inter1,inter2,mbranebh1,mbranebh2,mbranebh4,mscmc}; U-duality of
entropy of four and five dimensional black holes is discussed in {\em
e.g.} \cite{vijaycargese} and the comprehensive review \cite{dy}.

Counting the degeneracy of states for this system of open strings and
D-branes is simplest to perform in the limit \cite{anscv} where we
wrap D1,5-branes on a compactification manifold with
$R_5\!\gg\!V^{1/4}$, so we get a theory in $(1\!+\!1)$ dimensions; it
has ${\cal{N}}\!=\!(4,4)$ supersymmetry.  By analyzing the gauge
theory on the brane, one finds \cite{jmmthesis} that there are
$n_b\!=\!4Q_1Q_5$ bosonic and $n_f\!=\!n_b$ fermionic degrees of
freedom.  Roughly, these correspond to the open strings with one
endpoint on a D1 and one on a D5, and the 4 arises because the D1 can
move only inside the D5.  The central charge of this $d\!=\!1\!+\!1$
theory is thus $c\!=\!n_b\!+\!n_f/2\!=\!6Q_1Q_5$.  The more rigorous
way to calculate $c$ is via T-duality on $x^5$ where we get the moduli
space of instantons on $T^4$.  When the charges are macroscopically
large, this moduli space is well approximated \cite{bsv,vaf} by the
symmetric product space $S^{Q_1Q_5}(T^4)$, and its dimension is
$c\!=\!6Q_1Q_5$.  Now, the massless bosonic and fermionic degrees of
freedom carry left-moving momentum, and energy.  Since the
supergravity configuration is BPS, the configuration on the brane must
also be BPS.  Hence the energy and momentum in the $(1\!+\!1)$
dimensional theory are related by $E\!=\!P_L\!=\!Q_p/R_5$.  We can
then calculate the partition function for this system of bosons and
fermions:
\begin{equation}\label{zomega}
Z = \left[ \prod_{Q_p=1}^\infty {\frac{1+w^{Q_p}}{1-w^{Q_p}}}
\right]^{4Q_1Q_5} \equiv \sum \Omega(Q_p) w^{Q_p} \quad ,
\end{equation}
where $\Omega(Q_p)$ is the degeneracy of states at $d\!=\!1\!+\!1$
energy $E\!=\!Q_p/R_5$.  When $Q_p$ is large, the asymptotic behavior
of $\Omega(Q_p)$ is at leading order exponential,
$\Omega(Q_p)\!\sim\!\exp\sqrt{\pi{cEL_5}/3}$, with
$L_5\!=\!{2\pi}R_5$.  The microscopic entropy is therefore
\cite{anscv}
\begin{equation}\label{smicro}
S_{micro} = 2\pi \sqrt{Q_1 Q_5 Q_p} \quad ,
\end{equation}
which agrees precisely with the Bekenstein-Hawking entropy of the
black hole (\ref{5dbhentropy}).  This agreement between the entropy of
the D-brane/string boundstate and the black hole is thus in line with
expectations from supersymmetric nonrenormalisation arguments.

Let us now inspect the regime of validity of the D-brane picture
\cite{jmmans9609}.  We already have that the closed string coupling is
weak, $g\!\ll\!1$.  Feynman diagrams involving open strings pick up a
factor of $gQ_{1,5}$ because $g_{open}\!\sim\!\sqrt{g}$ and because of
the Chan-Paton factors on the open string endpoints.  As explained in
\cite{mrdjpans}, processes involving $Q_p$ give rise to factors $g^2
Q_p$ when propagators hook onto the external state.  Therefore,
conventional D-brane perturbation theory is good when
\cite{jmmans9609}
\begin{equation}\label{dbcorr}
r_{1,5,p} \ll \ell_s \quad ,
\end{equation}
which is precisely the opposite regime to (\ref{alphacorr}) where the
classical supergravity solution is good.  The D-brane/string
perturbation theory and black hole regimes are thus complementary.
This feature is related to open-closed string duality.  Note in this
regard a remark in \cite{mrdjpans} and the recent work \cite{juannew}
where it is argued that in the near-horizon region there is a duality
between the supergravity description and a large-N gauge theory
description.  We comment more on this in a Note Added at the very end
of this article.

A qualitatively different system has been studied which
gives rise to BPS $d=5$ black holes with nonzero entropy
\cite{mscmjp}.  There, the system of branes studied is wrapped
D5-branes with self-dual magnetic fluxes on the fivebranes.  The
degeneracy of states must be counted via the Landau degeneracy of open
strings connecting different branes, or alternatively using properties
of torons occurring in the supersymmetric gauge theories on the
branes.  Agreement is found with the Bekenstein-Hawking entropy.

Entropy of a more general BPS six dimensional black string with has
also been studied from the string theory perspective in
\cite{wavy1,wavy2,wavy3,wavy4,wavy5}.  In these works, left-moving
waves on the string \cite{dg92} are allowed to have large amplitudes.
The entropy can be calculated using both string theory and classical
gravity.  In this case, even though the classical spacetime is
singular on the horizon, agreement is again found between the two
different approaches.  The issue of the importance of quantum
corrections to the classical spacetime side of this agreement has not
yet been resolved fully.

\subsection{Fractionation}

An important subtlety arises in the use of the exponential
approximation to the formula (\ref{zomega}).  This approximation is
only valid when the energy of the system is such that there is a large
degeneracy of states at that given energy, or equivalently when
$Q_p\!\gg\!Q_{1,5}$.  As pointed out in \cite{jmmls}, the exponential
approximation is no longer valid when $Q_p\!\sim\!Q_{1,5}$.  The
simplest way to see this physically \cite{jmmls} is to recall that the
massless open strings constitute a gas of left-movers of order
$Q_1Q_5$ species with average energy $Q_p/R_5$.  Introducing a
temperature $T_L$ and making the assumption of extensivity we have
energy $E\!\sim\!Q_1Q_5R_5T_L^2\!=\!Q_p/R_5$ and
$S\!\sim\!Q_1Q_5R_5T_L$.  Eliminating $T_L$ gives
$S\!\sim\!(Q_1Q_5Q_p)^{1/2}$ which is the Bekenstein-Hawking scaling.
However, substituting back we find the inverse temperature
$T_L^{-1}\!\sim\!R_5(Q_1Q_5/Q_p)^{1/2}$, which if all three charges
are comparable is longer than the wavelength of a typical quantum in
the box of size $R_5$, and so the gas is too cold for thermodynamics
to be applicable.

The solution of this problem \cite{jmmls,jmmthesis} is to realize that
in this regime the intersection string is not multiple copies of
singly wound string, but rather a single copy of a multiply wound one.
This fractionation was suggested by the work of \cite{srdsdm9601};
fractionation may also be seen in the context of the $S^{Q_1Q_5}(T^4)$
theory.

Since it is the massless strings running between the D1,5 that
dominate the entropy of the configuration \cite{cgcjmm}, the
intersection string is wound \cite{jmmls} on a large radius
$R_5Q_1Q_5$ instead of $R_5$ and thus has an energy gap
$\delta{E}\!\sim\!1/(R_5Q_1Q_5)$ rather than the na{\"{\i}}ve $1/R_5$.
In addition, the tension of the intersection string becomes
fractionated by comparison to the na{\"{\i}}ve D-string tension.  Then
there are plenty of low energy modes available.  Using
$E\!\sim\!(R_5Q_1Q_5)T_L^2$, we find
$T_L^{-1}\!\sim\!(R_5Q_1Q_5)/(Q_1Q_5Q_p)^{1/2}$, so that the
temperature is plenty hot enough for the equation of state to be
valid.  The counting then proceeds in a similar manner as before, but
replacing $c=6Q_1Q_5$ by $c=6$ and $L_5$ by $L_5Q_1Q_5$.  The result
again agrees precisely with the Bekenstein-Hawking entropy.

\subsection{${\cal{N}}=2, d=4$ BPS Black Holes}

$d=4$ extensions of the IIB $d=5$ results were found in
\cite{cvjrrkrcm} by adding a new ingredient to the $\{R1,R5,W\}$
system: a KK6.  The T-dual system was studied in IIA
\cite{jmmans9603}.  U-dual $d=10$ configurations were found in
\cite{irkaat9604b,vbfl}, in more generality in \cite{vbrglfl}, 
and in \cite{irkaat9604b} an M-theory
configuration was also exhibited.  Picking one representative of the
duality orbit, we have
\begin{equation}
{\begin{array}{l}
M5_1 \\ M5_2 \\ M5_3 \\ MW
 \end{array}}
\left[ 
{\begin{array}{lllllllllll}
 0     & \cdot & \cdot & \cdot & 
 4     & 5     & 6     & 7     & \cdot& \cdot & \natural \\
 0     & \cdot & \cdot & \cdot & 
 \cdot & \cdot & 6     & 7     & 8     & 9     & \natural \\
 0     & \cdot & \cdot & \cdot & 
 4     & 5     & \cdot & \cdot & 8     & 9     & \natural \\
 0     & \cdot & \cdot & \cdot & 
 \cdot & \cdot & \cdot & \cdot & \cdot & \cdot & \natural
 \end{array}}
\right] \quad .
\end{equation}
Note that since M5-branes cannot end on any other branes
\cite{pkt9609}, this should really be thought of as fivebranes with
parts extending in different orthogonal directions.  The four
dimensional black hole is recovered by rolling up the above on a
manifold $T^6\times S^1$.  

This view of four dimensional black holes as composite M5-branes with
additional momentum can be extended to compactifications with less
supersymmetry.  There, instead of wrapping the fivebranes on pairs of
2-cycles in $T^6$ we wrap them on 2-cycles in a Calabi-Yau threefold
$CY_3$, to obtain BPS black holes in $d=4,{\cal{N}}=2$ supergravity.

Entropy formul\ae\ for the ${\cal{N}}\!=\!2$ black holes were obtained
in \cite{n20,n21,n22,n23}; in \cite{bcdklm} some quantum corrections
were studied.  For toroidal compactification a conjecture was made in
\cite{irkaat9604b} about how to count states in the microscopic
$1\!+\!1$ dimensional theory on the common string; in \cite{vbfl},
essentially the same observation was made independently, in the
$d\!=\!10$ context.  In \cite{dkdaljmmans} the microscopic entropy was
also computed heuristically for the ${\cal{N}}=2$ case for a
restricted class of solutions, and this was extended to the general
case in \cite{kbtm,jmmn2}.  See also the work \cite{sjrey}.

Precise methods for counting the microscopic entropy for BPS
${\cal{N}}\!=\!2,d\!=\!4$ black holes became available more recently
\cite{jmmansew}, and do not make use of string theory; related work in
the IIA context was done in \cite{cv9711}.  The microscopic
computations were done in a regime of parameter space where the
M5-branes' gravitational fields were negligible for the purpose of
counting states.  Certain charge restrictions were also implemented,
which meant that the fivebranes were smooth and noncoincident.  The
dimension of the space of supersymmetric M5-brane deformations, and
hence the degeneracy of states, is then calculable in the regime where
$R_\natural\!\gg\!V_{CY}^{1/6}$, by analyzing a ${\cal{N}}\!=\!(0,4),\
1\!+\!1$ dimensional theory obtained from knowledge of the chiral
worldvolume theory for the M5-brane and the Calabi-Yau $CY_3$.  The
microscopic entropy was then calculated, including the first
subleading term.

The charge restrictions were also needed in order that semiclassical
$d\!=\!11$ supergravity could be used for the strongly gravitating
side.  On the black hole side, the leading correction to the entropy
was computed in the Euclidean approach by evaluating the leading
correction in the $d\!=\!4$ supergravity action, which descended from
an $R^4$ correction to $d=11$ supergravity, on the classical solution.
In the IIA $d\!=\!10$ picture, this correction occurs at one string
loop, and it was argued that no higher string loops could correct it;
$\alpha^\prime$ corrections were also argued to be subleading due to
the charge restrictions.  The resulting black hole entropy and its
leading quantum correction then agree precisely with the M-brane
computation \cite{jmmansew}.

In this picture, we have seen that the microscopic degrees of freedom
behind the Be\-ken\-stein-Haw\-king entropy of the ${\cal{N}}\!=\!2,\
d\!=\!4$ black hole may be thought of as zero-cost deformations of the
M5-brane configuration, or ``M5-brane foam'' \cite{jmmansew}.  These
are analogous to the D-brane/string moduli space degrees of freedom in
the $d=1+1,\ (4,4)$ non-chiral theory from our previous example.  An
earlier remark about counting black hole entropy by considering
``D3-brane foam'' was contained in the work \cite{cgcjmm2}.

A recent calculation \cite{ass9712} has found qualitative agreement
between string state entropy and the sister black hole entropy for
${\cal{N}}=2,d=4$, in the same spirit as the calculation for maximal
supergravities \cite{ass}.  The undetermined constant for matching in
type II is shown to be different to that required for heterotic
matching, as expected.  Another recent work \cite{kbmcwas} has
considered near-BPS ${\cal{N}}\!=\!2$ black holes in the light of the
above precise M-brane counting.

We now leave BPS territory and turn to near-BPS configurations.

\subsection{Near-BPS $d=5$ Black Holes}

In ten dimensional string frame, the metric corresponding to the
nonextremal configuration with R-R 1- and 5-brane and momentum charge
is \cite{mcdy,gthjmmans}
\begin{eqnarray}
ds_{10}^2 & =& f_1(r)^{-1/2} f_5(r)^{-1/2} \left\{ dt^2 - dx_5^2 -
k(r)\left(\cosh\!\alpha_p dt + \sinh\!\alpha_p dx_5\right)^2
\right. \nonumber\\ & & \quad - \left. f_1(r)^{\mbox{}} \left[ dx_6^2
+ \ldots + dx_9^2 \right] - f_1(r) f_5(r)^{\mbox{}} \left[
\left\{1-k(r)\right\} dr^2 + r^2 d\Omega_3^2 \right] \right\} \quad ,
\end{eqnarray}
where $r^2 = \sum_{i=1}^4 (x^i)^2$ and 
\begin{equation}
k(r) = {\frac{r_H^2}{r^2}} \quad , \qquad f_{1,5}(r) = 1 +
k(r)\sinh^2\!\alpha_{1,5} \quad .
\end{equation}
Here the horizon radius $r_H$ is related to the ADM mass of the
pre-boost black hole.  Note that the BPS solution may be recovered by
taking the limit $r_H\!\rightarrow\!0, \alpha_i\!\rightarrow\!\infty$
in such a way that the product $r_H e^{\alpha_i}$ is finite.  To get a
$d\!=\!5$ black hole we roll up on the five-torus.  Then the form of
the metric and R-R antisymmetric tensor implies that there are four
independent length scales for the five dimensional black hole:
\begin{equation}\label{r15pH}
r_1^2 = {\frac{g Q_1\ell_s^6}{V}} \quad, \qquad
r_5^2 = g Q_5 \ell_s^2 \quad , \qquad
r_H^2 = {\frac{2}{\sinh\!\left(2\alpha_p\right)}} 
        {\frac{g^2Q_p\ell_s^8}{R_5^2 V}} \quad , \qquad
r_p^2 = \tanh\alpha_p {\frac{g^2 Q_p\ell_s^8}{R_5^2 V}} \quad .
\end{equation}
$r_{1,5}$ are the scales of variation of the supergravity fields due
to R1,5-brane charges.  In the BPS limit, $r_p$ is the scale of
variation due to the momentum charge.  

Arguments similar to those of the last section may be used to
calculate the D-brane degeneracy of states.  Near-BPS black holes
possess a small energy excess over their BPS counterparts, and they
break supersymmetry.  This added energy needs to be macroscopically
measurable but small by comparison to the energy of the BPS system.
We can add energy by adding pairs of strings moving with opposite
momenta, or by adding pairs of D1,5-branes and anti-D1,5-branes
\cite{cgcjmm}.

We work in the regime $R_5 \gg V^{1/4}$ where the theory is
effectively $(1\!+\!1)$ dimensional, so that $r_H, r_p \ll r_1, r_5$.
This means that the contributions to the entropy coming from
anti-D1,5-branes are suppressed by comparison to the contributions
from left- and right-moving open strings.  Then, since the string
coupling in five dimensions involves an inverse power of $R_5$ via
(\ref{Gd10}), interactions between left- and right-movers are weak and
we may use a dilute gas approximation \cite{gthans9602}.  In this
$(1\!+\!1)$ dimensional dilute gas regime, entropies of left and right
movers are additive.  {}From the point of view of the supergravity
solution, this is where the configuration is really a six dimensional
black string.

The lowest energy, and highest entropy, modes for the near-BPS system
are then the left- and right-moving open strings.  The left-movers for
the BPS system have a macroscopically large charge $Q_p\!=\!N_L$.  To
mock up the near-BPS black hole, on the D-brane side we add a few
left- and right-movers, $\delta N_{R,L}$, while keeping $\delta
N_R\!=\!\delta N_L$ in order to keep the charge
$Q_p\!=\!\left(N_L\!+\!\delta N_L\right)\!-\!\left(\delta N_R\right)$
fixed.  The microscopic entropy for the near-BPS configuration may
then be calculated by adding the left- and right-mover entropies,
using fractionation if the charge ratios warrant it.  The microscopic
entropy of near-BPS D-brane/string system then agrees precisely with
the Bekenstein-Hawking entropy,
\begin{equation}
S_{micro} = S_{BH} \quad .
\end{equation}

The case of the near-BPS plain D5-brane was analyzed in \cite{jmm5};
previous and subsequent work on BPS fivebranes may be found in
\cite{dvv5}.  It was found that the entropy can be accounted for in
terms of a gas of weakly interacting effective instanton strings with
fractionated tension.  It was, however, difficult to justify the
assumption that the gas is weakly interacting.  It would be
interesting to understand the reason for the precise agreement in this
picture.

Surprising agreement for extremal but non-BPS black holes
\cite{tmo9705} has also been found in
\cite{gthdaljmm,atish1,atish2,hjs}.

In recent work \cite{ksks}, the connection via previously unexplored
duality transformations \cite{sh04} between $d=4,5$ (and some
higher-$d$) non-extremal black holes and the $d=3$ BTZ \cite{btz,btz2}
black hole has been used to find the microscopic entropy.  The
degeneracy of states was computed using Carlip's method \cite{carlip},
which makes use of the fact that $d\!=\!2\!+\!1$ gravity can be
described by a Chern-Simons gauge theory.  In this way, intriguing
agreement with the Bekenstein-Hawking formula was found.  

\section{Emission and Absorption Rates}\label{emis}

Classically, black holes are known to gobble up particles crossing
their event horizons.  Semiclassically, they also Hawking radiate
\cite{swh}.  One way to calculate semiclassical emission rates is to
begin with the probability of absorption of a particular incoming
matter wave.  In the case of interest, we solve the wave equation for
the field in question, for low-energy modes.  In practice, solving
this equation can be very difficult technically because of nonlinear
mixing between different modes.  The wave equation has approximate
solutions where the radial coordinate $r$ is related to the
gravitational radii $r_i$ by $r \gg r_i$ (asymptotically far away) and
$r \ll r_i$ (near the horizon).  Matching of the approximate solutions
in the overlap region then yields the probability that a wave incident
from asymptotically far away will be absorbed by the black hole.  This
matching procedure can be complicated depending on the hierarchy of
scales of the different gravitational radii.  The absorption
probability for an ingoing wave differs from unity because the curved
geometry outside the horizon backscatters part of the incoming wave.
The dominant mode for either emission or absorption at low energy is
the $s$-wave; a general proof that the low-energy $s$-wave
cross-section for absorption of minimally coupled scalars by a $d$
dimensional spherically symmetric black hole is the area of the event
horizon was given in \cite{srdgwgsdm}.  The absorption cross-section
is then converted via detailed balance to an emission cross-section
and then an emission power using the canonical ensemble.

\subsection{Emission from D-Branes}

In \cite{cgcjmm} it was argued that Hawking radiation of near-BPS
black holes has an analogue in the D-brane system, and an approximate
rate for the process was calculated.  For clarity we will stay with
the five dimensional examples in the $\{ Q_1, Q_5, Q_p \}$ system, but
results for four dimensional black holes have also been obtained.  In
this D-brane system, there is a significant population of left-moving
excitations and a few right-moving excitations, all treated in the
dilute gas approximation.  The left-movers and right-movers carry a
net momentum and so different temperatures $T_{L,R}$ are assigned to
them.  In order for thermal equilibrium to be established, some
interactions are needed.  Now, when a left-moving and a right-moving
open string interact, they can scatter elastically or produce a closed
string which escapes to the bulk.  The latter process is the D-brane
analogue of the emission of a Hawking quantum.  The dynamics of the
D-brane/string system is calculable perturbatively at weak coupling,
and so the amplitude for this process may be computed precisely.  A
review of computation of closed-open string amplitudes relevant to
this emission may be found in \cite{ahirk}.

Since the number of right-movers is microscopically large, while being
macroscopically small, we may make the assumption that the system is
in thermal equilibrium, and that it is appropriate to use the
canonical ensemble.  If the number of right-movers were
microscopically small, the loss of one of them to make a bulk degree
of freedom would disrupt the ensemble too much and invalidate the
assumption.  We would also see trouble on the black hole side due to
the third law.  For a discussion of small corrections to the picture
we use, see \cite{peresko}.  Note also that thermalization of left-
and right-movers is indeed occurring, through decoherence processes
involving the bulk degrees of freedom.

{}From the partition functions for the left- and right-movers we can
then extract in the standard way the distribution functions for the
energies and hence the temperatures.  The distribution functions are
then the usual Bose-Einstein or Fermi-Dirac.  {}From this we can
proceed to find the emission rate for closed string quanta into the
bulk.  We need, as before, to be in the dilute gas regime which is
$r_{p,H}\!\ll\!r_{1,5}$, and \cite{jmmans9609}
$T_{L,R}^{-1}\!\gg\!r_{1,5}$; we also need low energies,
$\omega^{-1}\!\gg\!r_{1,5}$.  For the D-brane action we use the
Lagrangian for the effective intersection string, with fractionated
tension if needed.  In some cases, \cite{srdsdm9601,sdms96}, the
effective string picture has been checked explicitly by relating to
fundamental strings.

The first precise D-brane calculations \cite{srdsdm6} were made for
emission of bosonic scalars arising from internal components of the
ten dimensional graviton $h_{ij}$, which are minimally coupled
scalars.  It is simplest to use the harmonic gauge for the
perturbations.  The tree interaction vertex for these modes is then of
the form $h_{ij} \partial_\mu X^i \partial^\mu X^j$.  The next piece
of information needed is the field normalizations, which are obtained
from the brane and supergravity actions, remembering fractionation.
We can then find the amplitude for the two open strings to collide and
produce a bulk $h_{ij}$.  To find the emission power from the cross
section we average over initial states by using the thermal density of
states functions for the left- and right-movers, and sum over final
states.  The final answer for the energy emitted in an infinitesimal
energy range around $k_0$ per unit time is for the minimally coupled
scalars \cite{srdsdm6}
\begin{equation}\label{dbemiss}
{\frac{dE(k_0)}{dt}} = {\frac{A_H}{8\pi^2}}
{\frac{k_0^4dk_0}{e^{k_0/T_H}-1}} \quad ,
\end{equation}
where $A_H$ is the area of the event horizon.  The temperatures are
related to the Hawking temperature by \cite{jmmthesis}
$2T_H^{-1}\!=\!T_R^{-1}\!+\!T_L^{-1}$.  Absorption processes may be
calculated in an entirely similar fashion.  Four dimensional $s$-wave
results for minimally coupled scalars were obtained in \cite{db4}.
For a discussion of emission versus absorption in the D-brane context,
see \cite{adgmsrw}.  Note that the assumption of thermality for the
D-brane ensemble is crucial for reproducing the Hawking rate.

D-brane results in the $s$-wave then agree exactly with the
semiclassical black hole results, for emission and absorption of
minimally coupled scalars \cite{srdsdm6}, including all coefficients
and greybody factors.  As with the entropy, we are seeing whole black
hole functions reproduced from D-brane physics.  This is quite
remarkable, because the physical processes are different in the two
different approaches.

In string theory the low-energy effective theories are supergravities,
and these theories typically have nonminimal couplings between gauge
fields and scalars, so that old no-hair theorems do not apply
\cite{flfw}.  ``Fixed scalars'' \cite{n20,fixed2,fixed3} are scalar
fields in the black hole background experiencing a potential near the
horizon which fixes their values there.  An example of such a field in
our $d\!=\!5$ example is the dilaton; the emission rate for this was
found in \cite{srdsdm7}.  In \cite{dbfix2,dbfix3,dbfix4,dbfix5,ef2} it
was shown that emission of fixed scalars is suppressed by comparison
with minimal scalars, both from the supergravity and D-brane
perspectives.  On the D-brane side this was accomplished by showing
that processes involving fixed scalars arise from higher order
open-string processes on the brane worldvolume.  Precise agreements
were found for some situations.

Results have also been obtained for emission of higher angular
momentum modes \cite{dbang1,dbang2}, fermionic quanta
\cite{dbfer1,dbfer2,dbfer3}, charged particles
\cite{dbcharge1,dbcharge2,jmmans9609}, intermediate scalars
\cite{dbinter}, for rotating black holes \cite{dbrot,ef4}, and for
more general charge configurations \cite{srd9705,ef1,roberto,ef5}.
These additional cases, including fixed scalars, provide a more
stringent test of the effective string picture of near-BPS black hole
radiation than the initial calculations.  The case of plain D3-branes
\cite{d31} has also been studied extensively 
\cite{d32,d33,ssgirk97,ssgah98,sdmam,kh98} 
and some precise agreements have been found in that system as well.

\subsection{Discrepancies, and a Correspondence Principle Fix}

Discrepancies between the effective string model for $d\!=\!4,5$ and
semiclassical black hole calculations have arisen, in analysis of
higher dimension operators, higher partial wave emission, and black
holes with generic charge and angular momentum, {\em e.g.}  in
\cite{ef1,ef2,roberto,ef4,ef5}.  These discrepancies may be due to
insufficient understanding of the effective string model, and/or the
lack of an accompanying nonrenormalisation theorem.  In addition, in
\cite{sdms96} it was shown that the fractionation argument breaks down
for D1-branes in isolation and wrapped on $x^5$ when $R_5$ is too
small.  Assuming that this carries over for the D1-branes in the black
hole's sister boundstate, this means that $R_5$ should not be too
small in string units.  Otherwise, the excitation energy levels would
be smeared into broad resonances and the effective string model would
lose its calculational efficacy.  The case of parallel $p$-brane black
holes was discussed in \cite{srd9705}.

One situation in which there were emission disagreements
\cite{rels,roberto} appeared to clash with the correspondence
principle.  This was the case of a $d=5$ black hole with two large
NS-NS charges which, as we saw in a previous section, is sister to an
excited string state carrying winding and momentum charge.
Qualitative features of the emission spectra disagreed between the
string and black hole systems at the correspondence point, where they
should by rights agree.  A resolution of this problem was found in
\cite{sdm9706}.

The essential new input was the realisation that there are other
degrees of freedom than stringy excitations which can come into play
for the case of a $d=5$ black hole with charges $Q_w, Q_p$; namely,
NS5-branes wrapped around the compactification manifold.  Since
$M_{NS5} \sim 1/g^2$, these objects are much heavier than strings at
weak coupling and are therefore unimportant degrees of freedom.  As
the coupling is increased, however, the fivebranes become lighter.

Now, from (\ref{mns},\ref{qns}) we can see that the extra energy above
the BPS energy scales in $d=5$ approximately as
\begin{equation}
\Delta E \sim {\frac{r_H^2}{G_d}} \sim {\frac{R_5V r_H^2}{g^2
\ell_s^8}} \quad ,
\end{equation}
which at the correspondence point $r_H \sim \ell_s$ is \cite{sdm9706}
the mass of a single NS$5\!-\!{\overline{\mbox{NS}5}}$ pair.  However,
the fivebranes are fractionated by $1/(Q_w Q_p)$ in this regime, as
can be seen by U-duality from the $\{Q_1,Q_5,Q_p \}$ system.  To get
from there to our situation with $Q_w,Q_p,n_{NS5}$, we perform a
sequence of dualities, {\em e.g.}  $S \cdot T_5 \cdot S \cdot T_{6789}
\cdot S$.  In the process, additional left- and right-movers $\delta
N_R=\delta N_L$ become NS$5\!-\!{\overline{\mbox{NS}5}}$ pairs.
Therefore, there are actually many pairs of fractionated fivebranes
\cite{sdm9706}:
\begin{equation}
\left. n_{5{\overline{5}}} \right|_c \sim Q_p Q_w \quad ,
\end{equation}
and a corresponding entropy that is comparable \cite{sdm9706} to that
carried by strings, at the correspondence point:
\begin{equation}
\left. S_{NS5}\right|_c \sim \sqrt{Q_w Q_p} \sim
\left. S_{F1}\right|_c \quad .
\end{equation}
So we see that at the correspondence point the NS5-branes begin to
share the excess energy $\Delta E$.  At larger couplings the NS5-brane
entropy in fact dominates over that of the strings, while at weaker
couplings the string entropy dominates.

Now let us recall that gravity does not decouple for NS5-branes (when
$\ell_s$ is fixed).  Then the fact that the NS5-branes start to
contribute to the entropy just at the correspondence point ties in
with the fact that the correspondence point is just where the metric
deviates enough from flat space to be called a black hole
\cite{sdm9706}.  For four dimensional black holes, the extra branes
coming into play are the solitonic Kaluza-Klein monopoles
\cite{sdm9706}.

Then the entropy and the emission properties of the NS-NS black holes
can be shown to match at the correspondence point with those of the
$\{{\mbox{F1,NS5}}\!-\!{\overline{\mbox{NS5}}}\}$ system, with both
the F1- and NS5-branes controlling which polarizations may or may not
be emitted, just as the D1,5-branes and open strings running around
$R_5$ do for the U-dual R-R system \cite{sdm9706}.  General reasoning
of this sort is also suggestive of a way to resolve some difficulties
with emission and absorption of higher partial waves \cite{sdm9706}.

In addition, from (\ref{chargeneutral}) we saw that correspondence
points of two configurations related by a boost are mapped to each
other by the boost.  This can be used to relate emission rates in one
configuration to emission rates in the other
\cite{srdsdmskrpr,srdsdmpr}.

\subsection{Near-BPS Nonrenormalisation, and Information}

The agreement of the entropy and emission power for near-BPS black
holes and the sister D-brane/string states is so striking that we are
led to wonder if there is a nonrenormalisation theorem operating.  In
the BPS case there was a direct nonrenormalisation theorem available
to protect the degeneracy of states from quantum corrections as the
coupling is increased to take the sister D-brane/string boundstate
into a black hole.

For non-BPS configurations, nonrenormalisation theorems may indeed be
operating in order to ensure the precise agreement found for entropy
and emission for near-BPS black holes in the dilute gas regime.  For
the $d\!=\!5$ case in the dilute gas regime, an explanation was
offered in \cite{jmm9611}, where it was argued that at energies low by
comparison to the inverse gravitational radii $1/r_{1,5}$, the moduli
space description for the effective string is not renormalized
perturbatively or nonperturbatively.  This means that the entropy and
low-energy emission/absorption rates are the same for the black hole
and its weak-coupling sister D-brane/string boundstate.  For another
discussion relating to this system see \cite{sfhsrw}.

Another situation in which a nonrenormalisation argument for
low-energy modes in near-BPS backgrounds has been formulated
\cite{ssgirk97,irks97} is in the study of absorption and emission of
longitudinally polarized gravitons by threebranes.  There, the
conformal invariance of SYM in $(3+1)$ dimensions was used to argue
that the central charge, the quantity controlling the leading process
for emission and absorption of those low energy modes, is not
renormalized.

In addition, in \cite{srds97} it was argued that some one-loop open
string corrections to the tree level parallel D$p$-brane results for
particle emission are calculable and small at low energies.  {}From
this it is conjectured that the D-brane result may be extrapolated to
the black hole regime as long as the energies are kept lower than the
inverse of any gravitational radius in the problem.

We have seen in previous sections spectacular agreements between black
hole entropy and emission for near-BPS states.  This led some
optimists in the string theory community to speculate that string
theory eliminates the information problem for black holes, because
scattering involving D-branes and strings is unitary.  However, we
believe that this speculation may be premature.

Since the D-brane/string and black hole pictures are complementary, we
will obtain little direct insight into the black hole information
problem by using the unitary nature of perturbative D-brane/string
scattering.  We must instead tackle head-on the issue of probing a
black hole directly, in the string theory context.  The question of
whether this can be done is an issue of principle, of whether there is
a well-defined prescription for calculating.  Whether or not we can
subsequently perform the calculation is of less importance.

Among proposals that have been made, two of them \cite{tb9606,ans9606}
suggest that this probing of the internal state of the black hole can
indeed be done in principle, and they involve stringy hair on black
holes.  Some previous suggestions on how stringy hair, such as that
associated with the infinite number of gauge symmetries in string
theory, might be relevant to the black hole information problem may be
found in \cite{hair1,hair2,hair3,hair4,hair5}.  In \cite{ans9606},
with the assumption of the existence of a local low-energy effective
field theory to describe scattering, it was argued that the statistics
symmetry group for D-branes may be important for information
retrieval. There remain several outstanding issues, however, one of
which is that it is unclear whether just the above statistics symmetry
can be used to reconstruct the entire quantum state of the black hole.
In addition, it is unclear whether information about the black hole
may be recovered by using low-energy probes; for a discussion on this,
see \cite{tb9606}.  In different works \cite{jmmans9609,jmm9611} it
was argued that nonrenormalisation arguments for near-BPS cases in the
dilute gas regime may push the information loss problem to energies
higher than the inverse gravitational radii.  In \cite{rcmessay} it
was argued that decoherence may be an important means by which mixed
states arise.  Recent work on near-horizon geometry and string theory
such as \cite{juannew} may also be relevant to studying the
information problem.

The upshot is, we believe, that string theory with its current
technology has not yet resolved the information problem in black hole
physics, although it remains the most promising candidate theory to do
so.

\section{D-Probes and Matrix Theory}\label{dprobmatr}

In studies of scattering of strings off D-branes it was found
\cite{dscatt1,dscatt2,dscatt3,dscatt4,ahirk} that the strings can
probe D-branes down to distances only of order the string scale.  This
occurs essentially because the D-branes have a ``halo'' of open
strings around and ending on them, which then interact with the closed
strings in the bulk with a characteristic scale $\ell_s$.

\subsection{D-Branes as Probes}

D-branes themselves can also be used as probes
\cite{mrdgm,mrd9604,bds}.  The crucial new feature of using D-branes,
rather than strings, as probes, is that the distance scales probed by
D-branes depend on the string coupling $g$ \cite{shs} and are shorter
than $\ell_s$ at weak coupling.  The use of D-branes as probes of
black holes was first investigated in \cite{mrdjpans,mlem03,dvv04} and
later in \cite{jmm9705,jmp,chepaat,bisy}.

Two complementary approaches to the physics of D-brane probes exist.
The first is to use the action for a test D-brane in the supergravity
background, given by (\ref{sdbiwz}).  This is a valid scheme when
gravitational radii are large by comparison to the string scale.  The
second method is to use SYM perturbation theory for the D-brane
system, and this is valid in the opposite regime.  The two approaches
may be expected to agree if there is a supersymmetric
nonrenormalisation theorem operating.  Theorems for systems of
interest have not been proven generally; indeed, they are provably
unavailable in some situations; see \cite{mdns} regarding $p\!=\!2,3$.
Recent overviews of the considerable work done in comparing
D-brane(/Matrix) and supergravity forces on probes may be found in
\cite{chepaat,aats97,jmms97}.  We will only touch on these works here.

In order to evaluate the action (\ref{sdbiwz}) for a probe D$p$-brane
in a supergravity background, we need to specify the gauge, {\em i.e.}
the relation between the worldvolume coordinates $\sigma^\alpha$ and
spacetime coordinates $X^\mu$.  We choose an obvious generalisation of
the static gauge:
\begin{equation}\begin{array}{rlrl}
X^\alpha &= \sigma^\alpha \quad , \qquad 
         & \alpha&=0,1,\ldots p \quad ,\\
X^i      &= X^i(t)     \quad , \qquad 
         & i     &=(p+1),\ldots,9 \quad ,
\end{array}
\end{equation}
where the velocities $v^i\!=\!dX^i/dt$ are small.  We then pull
back the spacetime metric in $d\!=\!10$ string frame to the
$(p\!+\!1)$ dimensional worldvolume.  This yields a worldvolume metric
with time components different to those of the spacetime metric
because of the motion of the probe.

For our familiar $\{Q_1,Q_5,Q_p\}$ system, this yields for a test
D5-brane probe
\begin{equation}
S_5 = - \tau_{D5} \int d^{5+1}\sigma {\frac{1}{H_5(r)}} \left[
\sqrt{1-v^2H_1(r)H_5(r)H_p(r)} - 1 \right] \quad ,
\end{equation}
where $H_i\!=\!1\!+\!r_i^2/r^2$ are the familiar harmonic functions
describing the black hole background.  {}From the point of view of the
$d\!=\!5$ black hole, we are imagining that our probe is wrapped
around the appropriate compactification manifold, and so it looks like
a RR-charged particle.  Separating off the kinetic and potential
terms, we have for the supergravity potential at lowest order in
$v^2$,
\begin{equation}\label{v5sugra}
V_5^{\mbox{\footnotesize{SUGRA}}} = -{\tau_{D5}} {\frac{v^2}{2}}
\left[ H_p(r) H_1(r) -1 \right] \quad .
\end{equation}
In the moduli space approximation, when we keep only ${\cal{O}}(v^2)$
terms, we see that the probe does not see the gravitational field of
the constituents of the boundstate which are of the same type as
itself.  Also, the metric on the moduli space is easily read off from
the coefficient of $v^2$ in the action.  In \cite{mrdjpans} the case
of the D1-brane probe was analyzed with specific attention to the
geodesics on this moduli space.  Although the moduli space metric is
geodesically complete, there are geodesics that take probes into the
black hole.

Analogous probe actions can also be written for non-extremal
geometries \cite{jmm9705}.  Probing of extremal but non-BPS black
holes has also been studied in \cite{jmp,bisy}.

Now we turn to the computations on the D-brane side.  In this case, we
are considering a situation where the boundstate of a large number of
D-branes/strings is separated by a distance $r$ from the probe
D-brane.  Open strings stretching between the probe D-brane and the
D-branes in the boundstate appear as massive fields in the gauge
theory for the probe+boundstate.

For BPS situations these probe+boundstate gauge theories have been
analyzed comprehensively in \cite{chepaat,aats97}, whose notation we
follow here.  Let us consider the situation where we have $N$
D$p$-branes, possibly with other branes dissolved inside or
intersecting them.  Then, with the masses of the stretched strings
$m\!=\!r/\ell_s^2$ providing an infrared cutoff, the general
contribution to the $(p\!+\!1)$ dimensional gauge theory effective
action is, using power counting \cite{chepaat,aats97,jmms97},
\begin{equation}
\Gamma = \sum_{L=1}^\infty \left(g_{Y\!M}^2 N\right)^{L-1} \int
d^{p+1}\sigma \sum_n {\frac{c_{n,L} F^n}{m^{2n-4-L(p-3)}}} \quad ,
\end{equation}
where $c_{n,L}$ are constants.  We will compare directly the one-loop
term in this expansion with the leading supergravity dependence.  To
leading order, the term of interest is, at one loop, schematically
\cite{chepaat,aats97,jmms97}
\begin{equation}
{\hat{\Gamma}}^{(1)} = {\frac{c_{7-p}}{(2\pi)^p r^{7-p}}} \int
d^{p+1}\sigma \left\{ -{\frac{1}{8}} {\mbox{Str}}\left[ F^4
- {\frac{1}{4}}\left(F^2\right)^2 \right] \right\} \quad ,
\end{equation}
where $c_n = (2\pi)^n/(n\omega_{n+1})$.  There is also a two-loop
term, part of which is needed for the comparison with supergravity at
subleading order in $r$.

The way the calculation is done in the SYM theory is as follows
\cite{chepaat,aats97}.  The number of D5-branes is $N$; to represent
the D1-branes in the D5-brane worldvolume theory, we turn on self-dual
magnetic fields, and we excite waves to represent the last charge.  To
represent the relative motion of the probe and the boundstate, we turn
on a transverse scalar; this actually means we need to include an
extra term in ${\hat{\Gamma}}^{(1)}$ which is easily obtained by
T-duality from $p\!\rightarrow\!p\!+\!1$.  The precise coefficients
for all the gauge theory configurations are fixed by charge
normalizations.  Then the leading one-loop terms
${\hat{\Gamma}}^{(1)}$ may be computed.  The result is \cite{mrdjpans}
\begin{equation}
V_5^{\mbox{\footnotesize{SYM}}} = -\tau_{D5} {\frac{v^2}{2}} \left[
{\frac{r_1^2}{r^2}} + {\frac{r_p^2}{r^2}} \right] \quad ,
\end{equation}
which agrees with supergravity at lowest order.  The status of the
two-loop term in SYM is, as yet, less clear \cite{mrdjpans,chepaat}.

In all of these supergravity/SYM comparisons a subtlety arises,
concerning the dependence of the SYM result on the probe-type harmonic
function.  For precise agreement with SYM, it is found that the
probe-type harmonic function of the supergravity boundstate must be
altered as $H \rightarrow H-1$.  As pointed out in
\cite{chepaat,aats97,jmms97}, this can be thought of in two different
ways.  The first is to assume that the number of probe-type branes in
the boundstate is large by comparison to any other quantum numbers of
the boundstate, so that the $r$-dependent term in $H$ dominates.  The
second is to dualize the probe branes to D0-branes, and to think of
the D0-branes as coming from a null rather than spacelike reduction of
an M theory wave MW; this is related to the ``DLCQ'' prescription of
\cite{lsdlfq} which we will discuss shortly.

Other agreements between potentials for D-brane probes interacting
with charged backgrounds have been found; see \cite{chepaat,aats97}
for a recent overview with references.  In the $d\!=\!5$ near-BPS
black hole case, a new twist arises, namely that in order to obtain
agreement for the one-loop results we have to perform a change of
radial variable \cite{jmm9705}.  Note that since this is a coordinate
transformation it does not contain any physics, and that it could be
done term by term in the SYM loop expansion.

\subsection{Matrix Theory}\label{matrixtheory}

The low energy limit of M theory is given by eleven dimensional
supergravity.  Approximately a year ago the proposal was made
\cite{bfss} that M theory, the overarching theory of everything, is
described in the light front frame by a certain matrix model, known as
Matrix theory.  This proposal has passed some important tests, such as
reproducing some low-energy scattering amplitudes of eleven
dimensional supergravity, and the emergence of string perturbation
theory in compactification on a circle.  Matrix theory also has the
virtue that many dualities are derivable from the Hamiltonian.  Recent
reviews may be found in \cite{dvvs97,tommatrix,mbs97,biglenny}.  The
ideas of Matrix theory have also been applied to black hole systems.
Initially, Matrix theory analogues of D-brane results were discovered
for charged black holes \cite{mlem03,dvv04,mlem04}, but recent works
have made use of other intrinsic properties of Matrix theory to tackle
the case of neutral black holes
\cite{tbwfirkls1,irkls,gthem,ml,tbwfirkls2,hlaat,tbwfirk}.

The basic length scale associated with the dynamics of
D0-branes is \cite{kp,dkps}
\begin{equation}
\ell_{11} = g^{1/3} \ell_s \quad .
\end{equation}
For weak coupling $g$, this is considerably shorter than the string
scale.  As can be seen from (\ref{ell11}),
it is also the $d\!=\!11$ Planck length.

Now suppose that we wish to do physics with D0-branes at distances
$\ell\!\sim\!\ell_{11}$.  The dynamics is described via the open
strings which connect the D0-branes.  The mass of an open string
stretched between two D0-branes a distance $\ell_{11}$ apart is
$m\!\sim\!g^{1/3}/\ell_s$.  For weak coupling, this is an energy far
lower than that of the massive modes of the string.  Hence, if we wish
to study dynamics at $\ell\!\sim\!\ell_{11}\!\ll\!\ell_s$, we may
neglect massive open string modes and simply use the low-energy
effective gauge theory.  Closed string loop corrections are also
suppressed for weak bulk coupling $g$.  Next let us rescale the
energies and take weak bulk coupling \cite{kp,dkps} such that
\begin{equation}\label{mlim}
g\rightarrow 0 \quad, \qquad
E_f={\frac{g^{1/3}}{\ell_s}}={\frac{R_\natural}{\ell_{11}^2}} \quad
{\mbox{fixed}} \quad .
\end{equation}
Note that we are sending $\ell_s$ (and $\kappa_{10}$) to zero in this
limit.  Then the Lagrangian for the D0-branes is obtained by
dimensionally reducing $d\!=\!9\!+\!1$ SYM to $d\!=\!1\!+\!0$.  The
resulting Lagrangian is in string units ($A_0=0$)
\begin{equation}\label{ld0}
{\cal{L}}_0 = {\mbox{Tr}} \left\{
  {\frac{1}{2g}}{\frac{dX^i}{dt}}{\frac{dX^i}{dt}} 
+ {\frac{g}{4}}\left[X^i,X^j\right]^2 
+ \theta^T{\frac{d\theta}{dt}}
- \theta^T\gamma_i\left[\theta,X^i\right] \right\} \quad .
\end{equation}
The bosonic fields $X^i$ are $N\!\times\!N$ matrices and, in physical
situations where they commute, the diagonal entries describe the
positions in the $i$th direction of the $N$ D0-branes.  The fermionic
$\theta$ is the 16 component Majorana spinor superpartner of $X^i$.
If we then rescale all fields and time by appropriate powers of
$\ell_{11}$, \cite{kp,dkps} then we obtain a Hamiltonian in
$\ell_{11}$ units with the $g$-dependence as an overall factor:
\begin{equation}\label{hd0}
{\cal{H}}_0 = R_\natural {\mbox{Tr}}\left\{ {\frac{1}{2}}\Pi^i\Pi^i
-{\frac{1}{4}}\left[X^i,X^j\right]^2
+\theta^T\gamma_i\left[\theta,X^i\right] \right\} \quad ,
\end{equation}
where $\Pi^i$ are the canonical momenta for $X^i$.

The BFSS conjecture \cite{bfss} is that the large-$N$ limit of the
$d\!=\!10$ theory described via ${\cal{H}}_0$ describes $d\!=\!11$ M
theory in the light front frame.  It says that the fundamental degrees
of freedom of M theory are the nonrelativistic D0-branes, along with
the stretched unexcited strings that connect them.

For an indication of why the BFSS conjecture is reasonable, let us
study light front kinematics a little.  Consider an M-theory graviton
with $N$ units of longitudinal momentum, which in $d=10$ IIA language
is a bunch of $N$ D0-branes.  The $d=11$ graviton is massless, whereas
the $d=10$ D0-branes are massive and BPS.  Then
\begin{equation}
P_\natural = {\frac{N}{R_\natural}} \quad .
\end{equation}
Next, from the usual relation for massless particles in $d=11$,
$P_0\!=\!\sqrt{P_\perp^2+P_\natural^2}$, we see that
\begin{equation}
P_0=P_\natural+{\frac{P_\perp^2R_\natural}{2N}}+
{\cal{O}}\left({\frac{P_\perp^4}{N^2}}\right) \quad .
\end{equation}
Going to light-front coordinates, $x^\pm=x^0\!\pm\!x^\natural$, we
find that
\begin{equation}
P_-\simeq{\frac{N}{R_\natural}} \quad , \qquad 
P_+\simeq{\frac{P_\perp^2R_\natural}{2N}} \quad ,
\end{equation}
and the rough equality becomes exact in the $N\rightarrow\infty$
limit.  We can then see by studying ${\cal{L}}_0,{\cal{H}}_0$ that the
light-front energy $P_+$ for the bunch is precisely the energy given
by the D0-brane Hamiltonian ${\cal{H}}_0$ in (\ref{hd0}).  Energies
for other M-theory objects, such as M2- and M5-branes, may also be
obtained from the large-$N$ Matrix theory Hamiltonian, by arranging
the noncommuting matrices $X^i$ in such a fashion as to reproduce the
appropriate central charge in the supersymmetry algebra.

Matrix theory is a quantum mechanics with sixteen supersymmetries.
The fermionic superpartners $\theta$ encode spin information.  This
extended supersymmetry has remarkable consequences, one of which is
the enabling of the identification of asymptotic states such as the
graviton and its superpartners \cite{bfss}.  Supersymmetry is also
crucial in order for the {\em a priori} short-distance
($\ell\!\sim\!\ell_{11}$) Matrix theory calculations to reproduce
lowest-order long-distance graviton scattering in supergravity
($\ell\gg\ell_{11}$) \cite{bfss}.

Now, in order to study lower dimensional black holes, we need to know
how to do Matrix theory on compactified manifolds; for simplicity we
will concentrate on tori $T^p$.  Then the unexcited open strings
connecting the D0-branes may wind around the cycles of the torus, and
via T-duality this results in the theory of $N$ D$p$-branes, {\em i.e.}
$U(N)$ gauge theory in $(p+1)$ dimensions.  Using
(\ref{Rnatural},\ref{ell11}), we can express the usual T-duality
relation (\ref{tduality}) in terms of the quantities
$R_\natural,\ell_{11}$:
\begin{equation}\label{Rtilde}
{\tilde{R}}_i = {\frac{\ell_{11}^3}{R_iR_\natural}} \quad .
\end{equation}

Following the BFSS conjecture, the conjecture of \cite{lsdlfq} was
made that ${\cal{H}}_0$ for finite $N$ still describes M theory, but
in discrete light front\footnote{The light front is mistakenly,
although commonly, referred to as the light cone, so we will stick
with the common usage DLCQ to avoid confusion.}  quantization, DLCQ,
where we compactify on a lightlike circle.  Subsequently,
\cite{assmatrix} and \cite{nsmatrix} have provided an argument for the
relation between finite-$N$ Matrix theory and lightlike compactified M
theory.  The way this works is by taking compactification on the usual
spacelike circle of radius $R_\natural$, and boosting to a lightlike
circle, while holding certain quantities fixed.  The resulting theory
is called M$^\prime$ theory; by assuming eleven dimensional Lorentz
invariance, we have that M theory is then the same as M$^\prime$
theory.  To see how this goes, we will follow \cite{nsmatrix}.  We
boost up the spatial circle via rapidity $\delta$, where
\begin{equation}
\sinh\!\delta = {\frac{R_\star}{\sqrt{2}R_\natural}} \quad ,
\end{equation}
and take the infinite boost limit 
\begin{equation}
g \rightarrow 0 \quad , \qquad R_\star \quad {\mbox{fixed}} \quad ,
\end{equation}
so that the circle becomes null in the limit.  For a massive particle
carrying $N$ units of (integer quantized) longitudinal momentum, the
light front momenta become precisely
\begin{equation}\label{pppmprime}
P_-^\prime={\frac{N}{R_\star}} \quad , \qquad
P_+^\prime={\frac{\left(m^2+P_\perp^2\right)R_\star}{2N}} \quad .
\end{equation}
It is important that in this boosting procedure we hold fixed $E_f$
and the M-theory torus radii in $d\!=\!11$ Planck units
\cite{nsmatrix} 
\begin{equation}
E_f={\frac{R_\natural}{\ell_{11}^2}}=
E_f^\prime={\frac{R_\star}{(\ell_{11}^\prime)^2}} \quad , \qquad
\rho_i\equiv{\frac{R_i}{\ell_{11}}}=
\rho_i^\prime={\frac{R_i^\prime}{\ell_{11}^\prime}} \quad .
\end{equation}
We then get
\begin{equation}
g=\left({\frac{R_\natural}{\ell_{11}}}\right)^{\!3/2} =
R_\natural^{3/4} \left(E_f^\prime\right)^{\!3/4}\rightarrow{0}
\quad{,}\qquad
\ell_s=\left({\frac{\ell_{11}^3}{R_\natural}}\right)^{\!1/2}=
R_\natural^{1/4} \left(E_f^\prime\right)^{\!-3/4}\rightarrow{0}
\quad{,}
\end{equation}
as for the limit (\ref{mlim}) that produced the D0-brane Hamiltonian
(\ref{hd0}).  Going to the dual torus ${\tilde{T}}^p$ with volume
${\tilde{V}}_p=\prod_i {\tilde{R}}_i$, we get from (\ref{Rtilde}) the
dual radii
\begin{equation}\label{tildeRi}
{\tilde{R_i}}={\frac{1}{E_f\rho_i}}=
{\frac{1}{E_f^\prime\rho_i^\prime}}= {\mbox{finite}} \quad .
\end{equation}
Then the string coupling associated to the dual torus ${\tilde{T}}^p$
is, from (\ref{tduality}),
\begin{equation}\label{gtilde}
{\tilde{g}}=g {\frac{{\tilde{V}}_p}{\ell_s^p}}=
g{\frac{\ell_s^p}{V_p}}=
\ell_s^{3-p}\left(E_f^\prime\right)^3{\tilde{V}}_p \quad .
\end{equation}

Let us now inspect the coupling of the SYM theory on the dual torus.
It is dimensionful for $p\not =3$,
\begin{equation}\label{gtcplgs}
{\tilde{g}}_{Y\!M}^2 = (2\pi)^{p-2} {\tilde{g}} \ell_s^{p-3} \quad .
\end{equation}
{}From this we find a finite nonzero gauge coupling:
\begin{equation}\label{gymtilde}
{\frac{{\tilde{g}}_{Y\!M}^2}{(2\pi)^{p-2}}}=
{\frac{R_\natural^{3-p}\ell_{11}^{3(p-2)}}{V_p}}=
{\frac{R_\star^{3-p}(\ell_{11}^\prime)^{3(p-2)}}{V_p^\prime}}=
{\tilde{V}}_p\left(E_f^\prime\right)^3 = {\mbox{finite}} \quad .
\end{equation}

Now, as we can see from (\ref{gtilde}), the T-dual string coupling is
well-behaved for $p\leq 3$.  However, for $p>3$ it blows up in the
Matrix limit, so the bulk (supergravity) theory does not decouple from
the brane (gauge) theory, and we should instead use a dual description
\cite{t4,t5}.  In addition, the $d\!>\!3\!+\!1$ dimensional gauge
theory is strongly coupled in the ultraviolet, so new degrees of
freedom are required anyway, in order to make sense of the theory.
For $p\!=\!4,5$ \cite{nsmatrix,assmatrix} there is a well defined
prescription for the theory, at least at large-N \cite{jmas5b}.
However, for $p\!=\!6$ it is not clear whether the bulk theory can be
decoupled.  In the case of $p\!=\!4$, the appropriate theory is the
chiral theory of M5-branes compactified on a circle of finite radius,
with $\ell_{11}\rightarrow 0$ so that the gravitational field of the
M5-branes is negligible.  For $p\!=\!5$ an S-duality gives NS5-branes
with weak bulk coupling and finite worldvolume coupling; this may be
considered as a limit of a ``little string theory'' (see {\em e.g.}
\cite{ralh}).

At this point we make the comment that Matrix theory {\`{a}} la BFSS
is not yet a complete non-perturbative description of M-theory.  The
prescription of \cite{nsmatrix,assmatrix} does not produce agreement
between Matrix theory and supergravity scattering amplitudes in all
situations.  A difficulty appears to arise in compactifications on
some curved manifolds; a disagreement which has not yet been ironed
out was found for the K3 case by \cite{dougoo}.  (See however the
positive results of \cite{skales}.)  As we mentioned above, there are
also difficulties with compactifications on tori of dimension larger
than five, and these cases include toroidal compactification down to
the phenomenologically interesting case of four spacetime dimensions.
One apparent disagreement \cite{dineraj} seems, however, to have been
ironed out recently \cite{oyj}; see also \cite{wtivmvr,ffi,rejg}.
Another apparent objection \cite{hellp} has also been overcome and
even used to advantage in {\em e.g.} \cite{krogh}.  Lastly, it is not
known whether the BFSS conjecture is valid for any finite N (strong
conjecture), or whether the limit $N\rightarrow\infty$ is required
(weak conjecture).  Fortunately, the application of Matrix theory to
study the entropy of black holes of interest here does not rely on
these subtle details.

In the remainder of this section, we will first discuss the case of
charged black holes, and leave the qualitatively different case of
neutral black holes until later.  Matrix theory compactified on $T^p$
corresponds to $d\!=\!11$ M theory compactified on $S^1\!\times\!T^p$,
so for $d\!=\!5$ black holes we need $p\!=\!5$.  (For $d\!=\!4$ black
holes we would need $p\!=\!6$ which is not yet well enough
understood.)  For charged black holes, we will assume that the charge
represented by $N$ is large, and use SYM gauge theory as an
approximate low-energy effective Matrix theory.  This will be suitable
for the purposes of identifying the BPS states and their degeneracy.

\subsection{BPS Configurations}

Two different but related approaches have been used in studying $d=5$
BPS black holes in Matrix theory.  In both cases, agreement is found
with the Bekenstein-Hawking black hole entropy, and parallels with the
D-brane approach are seen.  We review the Matrix theory pictures of
\cite{mls97,dvv04} here.

The configuration in M theory which is dual to our familiar
$\{Q_1,Q_5,Q_p\}$ system is 
\begin{equation}
{\begin{array}{l}
M5 \\ MW \\ M2
 \end{array}}
\left[ 
{\begin{array}{lllllllllll}
 0     & \cdot & \cdot & \cdot & \cdot & 
 \cdot & 6     & 7     & 8     & 9     & \natural \\
 0     & \cdot & \cdot & \cdot & \cdot &
 \cdot & \cdot & \cdot & \cdot & \cdot & \natural \\
 0     & \cdot & \cdot & \cdot & \cdot &
 5     & \cdot & \cdot & \cdot & \cdot & \natural 
 \end{array}}
\right] \quad .
\end{equation}
The $d=11$ metric is \cite{mls97}
\begin{eqnarray}
ds_{11}^2 &=& H_{M2}(r)^{1/3}H_{M5}(r)^{2/3} \left\{
H_{M2}(r)^{-1}H_{M5}(r)^{-1} \left[dx^+dx^- - \left(H_{MW}(r)-1\right)
(dx^-)^2 \right] \right. \nonumber\\ & & \left. - H_{M2}(r)^{-1}
dx_5^2 - H_{M5}(r)^{-1} dx_{6\ldots 9}^2 - dr^2 - r^2d\Omega_3^2
\right\} \quad .
\end{eqnarray}
Compactifying $x^{5\ldots 9}$ on $T^5$ gives a black string stretched
along the $x^\natural$ direction, and upon compactification on
$x^\natural$ this yields our $d=5$ BPS black hole metric, with the
harmonic functions as before, with
$\{H_{M5},H_{MW},H_{M2}\}\!\rightarrow\!\{H_1,H_5,H_p\}$.  The
Bekenstein-Hawking entropy is the same calculated in any dimension and
we know the answer.

We now wish to compute the degeneracy of states in Matrix theory with
the same charges and mass as our supergravity solution.  Our theory is
$d=5+1$ SYM, on the dual torus ${\tilde{T}}^5$, and so various
spacetime charges will arise via electric and magnetic fluxes and
momentum along internal directions.  The dictionary relating M theory
quantities to those of the SYM theory on the dual torus is then
\cite{dvv04,mls97}
\begin{equation}\label{dictionary}
\begin{array}{llrl}
MW(i)            & \qquad\qquad & 
q_i              &= \int {\mbox{tr}} F_{0i} \\
M2(ij)           & \qquad\qquad & 
m_{ij}           &= \int {\mbox{tr}} F_{ij} \\
M2(i\natural)    & \qquad\qquad & 
m_{i\natural}    &= \int T_{0i} \\
M5(jklm\natural) & \qquad\qquad & 
f_i              &= \int {\mbox{tr}} \epsilon_{ijklm}F_{jk}F_{lm} 
\quad ,
\end{array}
\end{equation}
where $T_{\mu\nu}$ is the energy-momentum tensor in the SYM theory.
The case of an M5 wrapped on the $T^5$ does not have an analogue in
the effective $d=5+1$ SYM theory, because it would be a completely
delocalized state.

Our familiar configuration corresponds to turning on $m_{5\natural},
f_5$, and T-dualizing the five-torus.  Therefore, our $Q_1$ is the
number of M5-branes $f_5$, our D5-brane charge is $N$, and our $Q_p$
is the number of M2-branes $m_{5\natural}$.  Now, the M5-branes are
represented in the SYM theory by instanton strings.  These instanton
strings on the torus can be shown \cite{mls97} to have tension
fractionated by $1/N$.  They are therefore the important degrees of
freedom in the SYM theory and carry all of the entropy.  The light
front energy is\footnote{Here, $R_5,V$ are in fact the T-duals of our
familiar torus parameters.}
\begin{equation}
P_- = m_{5\natural} {\frac{R_\natural R_5}{\ell_{11}^3}} 
    + f_5 {\frac{R_\natural V}{\ell_{11}^6}}
    = M_{ADM} - {\frac{N}{R_\natural}}  \quad .
\end{equation}
{}From the dictionary (\ref{dictionary}) and from
(\ref{kappag},\ref{elld},\ref{Rnatural}) we see that the momentum and
winding charges on the instanton string are
\begin{eqnarray}
q_p &=& m_{5\natural} {\frac{R_\natural R_5}{\ell_{11}^3}} \quad , \\
q_w &=& f_5 {\frac{R_\natural   V}{\ell_{11}^6}} \quad .
\end{eqnarray}
The BPS instanton string has effective level number and entropy
\cite{mls97}
\begin{eqnarray}
N_L &=& {\frac{1}{2\pi\tau_{\mbox{\footnotesize{eff}}}}} \left[ P_-^2
- \left(q_p-q_w\right)^2 \right] \quad , \\ S &=& 2\pi\sqrt{N_L}
\quad ,
\end{eqnarray}
where $\tau_{\mbox{\footnotesize{eff}}}$ is the mass per unit length
of the instanton string.  We can calculate
$\tau_{\mbox{\footnotesize{eff}}}$ \cite{mls97} by using the usual
gauge theory formula for the mass of an instanton,
$4\pi^2/{\tilde{g}}_{Y\!M}^2$, and recalling that fractionation
occurs.  We then use the SYM coupling (\ref{gymtilde}) of the dual
torus to get
\begin{equation}
\tau_{\mbox{\footnotesize{eff}}} = {\frac{4\pi^2}{{\tilde{g}}_{Y\!M}^2
N}} = {\frac{VR_5R_\natural^2}{2\pi{N}\ell_{11}^9}} \quad .
\end{equation}
Putting it all together, we have for the Matrix theory entropy
\cite{mls97}
\begin{equation}
S_{\mbox{\footnotesize{matrix}}} = 2\pi\sqrt{f_5Nm_{5\natural}}
\quad{,}
\end{equation}
which is indeed the Bekenstein-Hawking entropy.  Note that
$R_\natural$ has disappeared from the entropy formula, as
required\footnote{If we imagined stretching our M2- and M5-branes
across the lightlike circle of radius $R_\star$ instead of the
spacelike circle of radius $R_\natural$, then the relation
(\ref{gymtilde}) ensures that we get the same entropy.}.

A related approach was studied in \cite{dvv04}, where fundamental IIA
strings naturally make an appearance; see also \cite{tomnati}.  In
this approach (the fivebranes are NS5's); we compactify Matrix theory
on a torus which has one small dimension, of radius $R_9$.  To make
contact with IIA strings, any other compact dimensions are taken to be
of order $\ell_s$, which is significantly larger.  Thus, on the dual
torus there is one large dimension only, and we can therefore
dimensionally reduce to $d\!=\!1\!+\!1$ SYM.  The SYM gauge coupling
turns out to scale as $1/R_9$, and so it is strong.  Then the SYM
theory gets stuck on the moduli space, while retaining the option for
some SYM fluxes that correspond, via a dictionary similar to
(\ref{dictionary}), to wrapped branes.  The entropy counting then
proceeds in a way similar to the D-brane counting, and the result
agrees precisely with the Bekenstein-Hawking formula.  A general
entropy formula, invariant under the duality group which remains upon
turning off the transverse fivebrane charge, was also derived via the
SYM theory in \cite{dvv04}.

Before we move on to neutral black holes, let us comment on the
near-BPS extension of the above.  Both of the approaches
\cite{mls97,dvv04} yield agreement with $S_{BH}$ if near-BPS
configurations are studied in the na{\"{\i}}ve $d\!=\!5\!+\!1$ SYM
theory.  In the approach of \cite{mls97}, it is argued that all of the
available degrees of freedom of the effective Matrix theory can carry
the entropy, not only those turned on in the BPS limit.  This is
somewhat different qualitatively from the D-brane picture of
\cite{cgcjmm}.  In addition, in studies of probing near-BPS
configurations with D0-branes (MW's), suggestions are made that a
better Matrix theory understanding of the interactions of the probe
with the boundstate may give clues useful for information retrieval
\cite{mlem03}.

\subsection{Neutral Configurations}

We saw above that we needed to have large-$N$ in order to be able to
use SYM gauge theory as our low-energy effective Matrix theory.  If
there is no large charge on a black hole, as is the case for {\em
e.g.}  Schwarzschild-type solutions, then we need a different
approach.  This case was the subject of
\cite{tbwfirkls1,irkls,gthem,ml,tbwfirkls2}, and we will discuss it
here; see also \cite{biglenny}.  For these neutral black holes, the
precise coefficient in front of the entropy is not available, but
scaling agreement can be found.  We will therefore drop all numerical
constants; for example, the distinction between the spacelike radius
and the lightlike one in DLCQ.

The strategy employed for neutral black holes is to boost the black
hole in the longitudinal direction, in order to turn on a significant
longitudinal momentum.  The idea is then to use Matrix theory degrees
of freedom to count the entropy.  Suppose we begin with a black hole
of mass $M$.  Under the boost, the longitudinal momentum becomes
\begin{equation}\label{pminusM}
P_- \sim {\frac{N}{R_\star}} = {\frac{N}{MR_\star}} M \quad .
\end{equation}
The light front energy is from (\ref{pppmprime})
\begin{equation}\label{pplusM}
P_+ \sim {\frac{MR_\star}{N}} M = {\frac{M^2R_\star}{N}} \quad .
\end{equation}
Here $N$ is large but finite.

The object of interest is a $d\!=\!(11\!-\!p)$ dimensional neutral
black hole, where one direction is compactified, namely the
longitudinal direction of M theory.  Then the appropriate solution is
an infinite array of $d$ dimensional neutral black holes; let us space
the members of the array in the eleventh dimension by
$\sim\!{\hat{R}}_\star$.  When $r_H/{\hat{R}}_\star\ll 1$, this array
solution is stable.  However, when $r_H/{\hat{R}}_\star\!\sim\!1$, the
array becomes unstable and turns into a black string stretched along
the longitudinal dimension.  When we boost the neutral array or
product, the transition point changes \cite{gthem,srdsdmskrpr}.

First, let us consider the product solution, {\em i.e.} the black
string.  Let us see what happens to the metric in the boost
directions.  The longitudinal piece of the metric is, with
${\hat{z}}\equiv {\hat{x}}^\natural$,
\begin{eqnarray}
ds_\parallel^2 &=&
\left(1-{\frac{\eta_H^{d-4}}{\eta^{d-4}}}\right)d{\hat{t}}^2
-d{\hat{z}}^2= d{\hat{t}}^2-d{\hat{z}}^2
-{\frac{\eta_H^{d-4}}{\eta^{d-4}}}d{\hat{t}}^2 \nonumber\\
&=&dt^2-dz^2 - {\frac{\eta_H^{d-4}}{\eta^{d-4}}} \left[\cosh\!\gamma_p
dt+\sinh\!\gamma_p dz\right]^2 \nonumber\\&=&
dt^2\left[1-{\frac{\eta_H^{d-4}}{\eta^{d-4}}}\cosh^2\!\gamma_p\right]-
dz^2\left[1+{\frac{\eta_H^{d-4}}{\eta^{d-4}}}\sinh^2\!\gamma_p\right]- 
dtdz\sinh\!2\gamma_p{\frac{\eta_H^{d-4}}{\eta^{d-4}}} \quad ,
\end{eqnarray}
where $\eta_H$ is the radius of the $(d\!-\!1)$ dimensional black hole
we used to make the black string.  {}From the metric, we see
immediately that if the radius of the $z$ dimension is $R_\star$ at
infinity, then at the horizon the radius is
\begin{equation}\label{rprinf}
R_p = \cosh\!\gamma_p R_\star \sim e^{\gamma_p} R_\star \quad ,
\end{equation}
and so the longitudinal dimension becomes large at the horizon of the
boosted configuration.  By direct analogy with
(\ref{mrr},\ref{qrr},\ref{sbhrr}), we have at large boosts $\gamma_p$
the black string parameters
\begin{equation}\label{epsp}
E_p \sim P_p \sim {\frac{V_pR_\star\eta_H^{d-4}}{G_{11}}}
e^{2\gamma_p} \quad ,\qquad
S_p\sim{\frac{V_pR_\star\eta_H^{d-3}}{G_{11}}}e^{\gamma_p} \quad .
\end{equation}
The product solution becomes unstable to forming an array when
$\eta_H\!\sim\!R_p$.  At this transition point, from
(\ref{rprinf},\ref{epsp},\ref{pminusM}) we find that the product has
entropy \cite{gthem}
\begin{equation}
\left.S_{p}\right|_t \sim \left.P_{p}\right|_t R_\star \sim N \quad .
\end{equation}

Next, let us boost the array.  Let the asymptotic spacing be, before
the boost, ${\hat{R}}_\star=R_\star\cosh\!\gamma_a$.  This ensures
that the spacing after the boost is $R_\star$.  Then the energy and
momentum after the boost are 
\begin{equation}
E\sim M\cosh\!\gamma_a \sim M e^{\gamma_a} \quad , \qquad P\sim
M\sinh\!\gamma_a \sim M e^{\gamma_a} \quad ,
\end{equation}
while the entropy is invariant,
\begin{equation}
S_a \sim {\frac{V_pr_H^{d-2}}{G_{11}}} \quad .
\end{equation}
The boost also causes \cite{gthem} an enlargement of the longitudinal
radius at the horizon of the black holes in the array.  The
instability to forming a product solution then sets in at
\begin{equation}
r_H \sim e^{\gamma_a} R_\star \equiv R_a \quad .
\end{equation}
At this transition point, we find that the entropy of the array is
\cite{gthem}
\begin{equation}\label{strans}
\left.S_{a}\right|_t \sim \left.P_{a}\right|_t R_\star \sim N \sim M
r_H\quad .
\end{equation}
Thus we see that at transition both the supergravity array and product
have entropy of order $N$, the number of units of longitudinal
momentum.  The array will dominate for larger $N$, while the product
will dominate for smaller $N$ \cite{gthem}.  

It was argued in \cite{tbwfirkls1} that in order to count Matrix
degrees of freedom, there is an optimal value of $N$.  The logic goes
that if $N$ is too small, we will not have enough Matrix degrees of
freedom to account for the entropy of the neutral black hole, while if
$N$ is too large, we will have too many degrees of freedom and most of
them will be frozen into their groundstate in a complicated way.  That
optimal value is argued to be that given by the common entropy of the
array and product at transition \cite{tbwfirkls1}.  We now turn to
examining the nature of the Matrix theory degrees of freedom, near
transition.

First let us consider the product side of the transition.  The
appropriate M theory degrees of freedom are stretched along the
longitudinal direction, like the black string.  These can only be M2-
or M5-branes.  The first, and cleanest, Matrix theory calculation for
neutral black holes was done \cite{tbwfirkls1} for $d=8$, for which we
do (DLCQ) Matrix theory on ${\tilde{T}}^3$.  Then the modes relevant
to the discussion are M2-branes, and they have energy
\begin{equation}
E_{M2} \sim {\frac{R_\star R_i}{\ell_{11}^3}} \sim
{\frac{1}{{\tilde{R}}_i}} \quad .
\end{equation}
These modes therefore correspond via (\ref{dictionary}) to gluons on
the dual torus, with momentum in the $x^i$ direction.  We can find the
entropy of these left- and right-moving gluons by using the conformal
invariance of $3+1$ dimensional SYM theory to write an approximate
equation of state and the entropy \cite{tbwfirkls1}.  Since this SYM
is the theory of $N$ coincident D3-branes, we have ${\cal{O}}(N^2)$
degrees of freedom, and
\begin{equation}\label{e3s3}
E_3 \sim N^2 {\tilde{V}}_3 T^4 \quad ,\qquad
S_3 \sim N^2 {\tilde{V}}_3 T^3 \quad .
\end{equation}
Now, from the T-duality relation (\ref{Rtilde}), the light front
energy (\ref{pplusM}) and the fact that $S\!\sim\!N$ around
transition, we find upon eliminating the temperature $T$
\cite{tbwfirkls1}
\begin{equation}
\left.S_3\right|_t \sim M (G_8 M)^{1/5} \sim M r_H \quad ,
\end{equation}
which is the entropy of a $d=8$ black hole at transition
(\ref{strans}).  Therefore, the Matrix theory degrees of freedom have
reproduced the entropy of the black hole.  We can now go back to the
equation of state to check on the temperature at transition.
Substituting back, we find
\begin{equation}
T_t \sim \left(N{\tilde{V}}_3\right)^{-1/3} \quad ,
\end{equation}
which is much smaller than the inverse size of the box, and so it
seems too cold for the above equation of state to be applicable.
However, we have forgotten about fractionation which takes place at
low temperatures for wrapped D-branes.  At low temperature, the
fractionated D3-brane can support fractionated momenta, quantized in
units of $1/(n_i{\tilde{R}}_i)$, where $n_i$ is the winding in the
$i$th direction.  The volume of the fractionated brane is the same as
for $N=n_1n_2n_3$ singly wound branes.  For the fractionated brane,
the above equation of state is then still valid as long as
$T\!>\!1/(n_i{\tilde{R}}_i)$.  Therefore, the lowest temperature at
which we may apply our equation of state is approximately
\cite{tbwfirkls1}
\begin{equation}
T\sim\left(n_1n_2n_3{\tilde{V}}_3\right)^{-1/3} \quad ,
\end{equation}
which is precisely $T_t$.  Thus we are right on the edge of validity
of (\ref{e3s3}).  Transforming this transition temperature back to the
unboosted frame gives the Hawking temperature of the original neutral
black hole \cite{tbwfirkls1}.

In \cite{irkls} similar reasoning was used to extend the discussion to
other $p$.  There, somewhat peculiar equations of state were required
for the SYM gluon gas in order to reproduce the black hole entropy at
transitions for $p\not =3$.  Those equations of state were, however,
motivated from studies of parallel $p$-brane black holes, such as
\cite{irkaat9604}.

Now we turn to the array side of the transition.  When the SYM gas of
gluons becomes too cold, which is the case for $N$ larger than its
transition value, the gluons freeze out and we are left to do zero
mode physics.  These zero modes are the motions of the D0-branes
themselves.  In general, the nature of the boundstate is very
complicated, but at $N$ near its transition value there is a simple
model \cite{tbwfirkls2,gthem} which succeeds in reproducing the
Bekenstein-Hawking scaling.  (For an earlier idea in this direction
see \cite{volovich}.)

The effective Lagrangian for the D0-branes was found at one loop SYM
in \cite{dkps} (to two loops in \cite{bbpt}).  {}From the supergravity
point of view it is found by letting a probe move in the background of
the other D0-branes.  To one loop the effective D0-brane Lagrangian
is, up to constants of order unity,
\begin{equation}
{\cal{L}}_{0,{\mbox{\footnotesize{eff}}}} \sim {\frac{Nv^2}{R_\star}}
+ {\frac{N^2G_{11}v^4}{R_\star^3V_pr^{d-4}}} \quad ,
\end{equation}
Note that the potential term is a gravitational attraction, which
depends on the velocities of the D0-branes.  It is therefore
reasonable to assume that the bunch of D0-branes forms a boundstate,
with the binding energy coming from the gravitational attraction.

The method of \cite{tbwfirkls2,gthem} is to use a mean-field approach,
and to assume that each D0-brane is in the boundstate of size $r_b$
and saturates the uncertainty bound,
\begin{equation}\label{heis}
{\frac{vr_b}{R_\star}} \sim 1 \quad .
\end{equation}
Because our D0-branes must be nonrelativistic, we need
$r_b/R_\star\!\gg\!1$.  Then, assuming spherical symmetry and that any
spin-dependent forces average to zero, applying the virial theorem
then gives a relation between the boundstate size and $N$.  Our light
front relation (\ref{pplusM}) then yields
\begin{equation}
M \sim {\frac{r_b^{d-3}}{G_d}} \quad ,
\end{equation}
which reproduces the black hole mass if the boundstate is the same
size as the black hole \cite{tbwfirkls2,gthem}
\begin{equation}
r_H \sim r_b \quad .
\end{equation}
{}From (\ref{pminusM},\ref{heis}) and the entropy at transition, we
then see that the D0-branes are indeed nonrelativistic at transition.
The next step is to assume that the bunch of D0-branes is a Boltzmann
gas, so that they are distinguishable.  Then the entropy is extensive,
of order $N$, and is \cite{tbwfirkls2,gthem}
\begin{equation}
S \sim N \sim r_H M \quad .
\end{equation}

This simple mean-field model is therefore remarkably successful in
reproducing the entropy for the $d$ dimensional, longitudinally
compactified, black hole at transition.  It is also argued in
\cite{tbwfirkls2} that emitting a D0-brane corresponds to Hawking
radiation.  In the very recent work \cite{tbwfirk}, a calculation of
the rate of this emission was made, and scaling agreement was found
with the Hawking emission rate for the sister black hole.

The assumption of distinguishability, however, is difficult to
justify.  Certainly, for normal D0-branes, we have either
Bose-Einstein or Fermi-Dirac statistics, depending on whether any spin
is turned on.  In a recent work, a model has been proposed
\cite{tbwfirkls2} in which the distinguishability assumption is
satisfied.  The analysis of \cite{tbwfirkls2} depended on the
dimension $d$.  The strategy used there was to turn a matrix theory
background which does not contribute significantly to the entropy but
breaks the D0-brane statistics symmetry and ``tethers'' each D0-brane
to the background in such a way that the D0-branes are
distinguishable.

In \cite{hlaat} an all-loop corrected ${\cal{L}}_0$ was used to
analyze the boundstate of D0-branes, and to study probe interactions
with it.  The resulting picture is somewhat different but again gives
an entropy in a certain regime which agrees with that of the array at
transition.  D0-probe physics was also studied in \cite{ejmli98}.  In
the recent work \cite{dkgl} an interesting proposal was made that a
tachyon instability may signal the existence of the horizon; it would
be interesting to find further evidence for this idea.  In
\cite{dal98} it was proposed that the entropy counting problem in 
Matrix theory could be mapped onto the original Gibbons-Hawking
calculation, assuming eleven dimensional Lorentz invariance of Matrix
theory.  Proof of $d\!=\!11$ Lorentz invariance of Matrix theory is
however an extremely subtle unsolved problem.

In summary, we see that in regimes where we can calculate in Matrix
theory, we reproduce black hole entropy successfully.  There remain
many interesting issues in the Matrix approach to black holes. 

\section*{Outlook}\label{outack}

We have seen that superstring theory has successfully given an
identification of the degrees of freedom giving rise to the
Bekenstein-Hawking entropy for black holes.  In general situations,
the entropy of the strongly gravitating system and that of its weakly
gravitating sister system agree qualitatively at the correspondence
point, while for BPS and near-BPS systems there are precise agreements
for the entropy and low-energy emission/absorption processes over a
wider region of parameter space.  We have also seen the Matrix theory
conjecture reproduces black hole entropy successfully for some cases.

Perhaps the most interesting feature of the stringy approach to black
hole entropy in diverse dimensions is that it encompasses many
different calculational viewpoints, some consistently related by
dualities to one another.  We have seen various methods for computing
the microscopic entropy, operating in dimensions all the way from
eleven to three.  In all cases, M-/D-branes and closed/open
fundamental strings, have played a crucial r{\^{o}}le.

String/M theory is still a work in progress, and there are several
outstanding issues to resolve from studies to date of black hole
systems in string theory.  One of the most important is the issue of
information loss, and how it relates to various microscopic entropy
computations.  It is our expectation that further progress in string
theory will lead to better calculational tools for addressing issues
in black hole physics and quantum gravity.  We look forward to an
exciting future in this area of study.

\subsection*{Acknowledgements}

The author wishes to thank Aki Hashimoto, David Gross, Gary Horowitz,
Rob Myers and Joe Polchinski for discussions, and especially Simon
Ross for discussions and reading over v1 of the manuscript.

\bigskip

\noindent This work was supported in part by NSF grant PHY-94-07194.
 
\section{Note Added}

The full import of the paper \cite{juannew} by Maldacena had not been
appreciated by the string theory community at the time this article
was written.  However, this changed early in 1998 when the focus
shifted away from Matrix Theory to what has become known as the
``AdS/CFT Correspondence''.  Here we must be very brief, and so we
give only a flavor of this topic without attempting to be
comprehensive.  A more in-depth appreciation may be gained by
consulting {\em e.g.} the {\it Strings '98} Online Conference
Proceedings \cite{strings98}.

\subsection{The AdS/CFT Correspondence}

We noted during the discussion of BPS black holes in subsection
\ref{bpsd5} that the usual perturbative gauge theory and black hole
regimes are complementary: they have no common region of validity.  We
also discussed the decoupling limit in subsection \ref{matrixtheory}
where we gave a brief overview of Matrix Theory.  In this decoupling
limit, although the coupling of the gauge theory on the branes remains
finite and nonzero, the coupling between the gauge theory on the
branes and the string theory in the bulk is turned off, as is the
gravitational field of the branes.  The crucial observation made by
Maldacena was that the decoupling limit can actually yield a system
with a nontrivial gravitational field if in addition one sends the
number of branes $N$ to infinity.  This reveals a new duality between
the large-$N$ limit of a gauge theory and a strongly gravitating
system.

In the original paper of \cite{juannew}, the brane systems studied
were those with constant (or no) dilaton: the D3 (and M2 and M5), and
the D1+D5.  In the large-$N$ decoupling limit (an analog of which is
readily available in $d\!=\!11$) the gravitational influence of the
branes is retained, but is frozen out because fluctuations become
prohibitively expensive in the limit.  In addition, the decoupling
limit zooms in on the near-horizon region of the brane geometry.  For
the D3, this region is $AdS_5\times S^5$.  To see this, let us examine
the D3 metric:
\begin{equation}
ds^2 = {\frac{1}{\sqrt{H_3}}} \left(dt^2 -
d{\vec{x}}_\parallel^2\right) -
\sqrt{H_3} \left(d{\vec{x}}_\perp^2\right) \quad ,
\end{equation}
where, with $r=\left|{\vec{x}}_\perp\right|$, 
\begin{equation}
H_3 = 1 + {\frac{4\pi g N\left(\alpha^\prime\right)^2}{r^4}} \quad .
\end{equation}
The large-$N$ decoupling limit may be expressed as
\begin{equation}
\alpha^\prime \rightarrow 0 \quad , \qquad
N\rightarrow\infty \quad , \qquad
g_{YM}^2 \quad {\mbox{finite}} \quad .
\end{equation}
To keep interesting dynamics in the gauge theory we must also keep the
masses of strings stretched between branes fixed:
\begin{equation}
U \equiv {\frac{r}{\alpha^\prime}} \quad {\mbox{finite}} \quad .
\end{equation}
In taking the decoupling limit we see that we lose the 1 in $H_3$, and
zoom in on the near-horizon region of the D3 geometry.  The length
scale set by the geometry, the supergravity radius, is
\begin{equation}\label{3bsgr}
r_3^4 \equiv 4\pi g N \left(\alpha^\prime\right)^2 \quad .
\end{equation}
Then with a change of radial coordinate $z \equiv r_3^2/U$, the D3
geometry becomes in the decoupling limit
\begin{equation}
ds^2 \rightarrow {\frac{r_3^2}{z^2}} \left(dt^2 -
d{\vec{x}}_\parallel^2 - dz^2\right) - r_3^2 d\Omega_5^2 \quad .
\end{equation}
There is also a R-R flux threading the sphere.  Thus on the
``supergravity [string] side'' of this duality we study string theory
on $AdS_5\times S^5$, a geometry with isometry group $SO(4,2)\times
SO(6)$.  Via standard Kaluza-Klein reduction, this can be viewed as a
theory on just the five-dimensional $AdS_5$.  On the ``gauge theory
side'' of the duality, one has ${\cal N}=4$ supersymmetric gauge
theory in four dimensions, which is known to be a conformal field
theory, at large-$N$.  The $SO(4,2)$ appears as the conformal group
and the $SO(6)$ as the R-symmetry group of the gauge theory.
Maldacena's conjecture is that these two theories are dual to one
another and is known as the ``AdS/CFT correspondence''.

As with Matrix Theory, there are weak and strong forms of the
conjecture.  Defining the 't Hooft coupling $\lambda^2 \equiv g_{YM}^2
N$, we find using (\ref{3bsgr}) and the $p\!=\!3$ relation
$g_{YM}^2=2\pi g$
\begin{equation}
{\frac{\alpha^\prime}{r_3^2}} =
{\frac{1}{\sqrt{2}}}{\frac{1}{\lambda}}
\quad ,\qquad 
g = {\frac{\lambda^2}{2\pi}} {\frac{1}{N}} \quad .
\end{equation}
At fixed but large $\lambda,N$ we see that the
strong['t~Hooft]-coupling expansion of the gauge theory corresponds to
the $\alpha^\prime$ expansion in string theory, and the $1/N$
expansion to the string loop expansion (see e.g. \cite{grossoo}).  The
strongest conjecture states that the duality holds order by order in
$1/\lambda$ and in $1/N$, while weaker conjectures state that it holds
only as $\lambda\!\rightarrow\!\infty$ and/or
$N\!\rightarrow\!\infty$.  We see that the AdS/CFT correspondence, if
correct, provides us in this D3 case with a nonperturbative definition
of string theory on $AdS_5\times S^5$.  It also says that the
large-$N$ master field of the gauge theory is provided by
supergravity.

Subsequent work has generalized the conjecture in many ways.  One
generalization \cite{iljuan} handled other BPS D$p$-brane systems, an
example of which is a large collection of D0-branes.  It can then be
seen \cite{shenker} that Matrix Theory is an example of ``Maldacena
duality''.  The original AdS/CFT conjecture was refined significantly
by \cite{gkp} (following earlier studies of absorption/emission of
bulk modes by D-brane systems), and independently by \cite{witten1}.
There, prescriptions were given for computing correlation functions in
the large-$N$ gauge theory by using supergravity vertices.  Also, it
was proposed that the place where the gauge theory lives can be
thought of as the boundary at infinity of the $AdS_5$ space.  This
highlights one of the surprising features of the new duality, namely
that of ``holography'': the physics of the (KK-reduced) gravitational
theory on $AdS_5$ is given by the (dual) large-$N$ gauge theory in
four dimensions.  For more on holography see
\cite{witten1,edlenny,jlfber} and talks in \cite{strings98}.
Subsequent computations of the BPS spectra, anomalies, etc.  on both
sides of the duality have provided additional evidence for the AdS/CFT
conjecture.  Another generalization \cite{witten2} covers the non-BPS
case, which is conjectured to give finite-temperature gauge theory.
Computations have been performed of everything from Wilson loops to
four-point correlations functions to glueball masses, assuming the
validity of the AdS/CFT duality conjecture.

So far, more has been learned about gauge theory by using supergravity
than the other way around, but we are optimistic that work in the
other direction will prove fruitful as well.

\subsection{Black Hole Entropy and AdS/CFT}

New methods for computing the Bekenstein-Hawking entropy of black
holes have been investigated since the AdS/CFT correspondence
\cite{juannew} came to light.  

We have seen that five-dimensional black holes can be thought of as
six-dimensional black strings compactified on a circle.  The
near-horizon geometry of the black strings turns out to be
$AdS_3\times S^3$.  More precisely, the first factor is a BTZ black
hole, which is everywhere locally (but not globally) $AdS_3$.  A new
counting of black hole entropy was done in \cite{andynew}.  (Although
the counting was probably motivated by the stringy AdS/CFT
correspondence, the method used was actually independent of it.)
String theory on $AdS_3$ is dual to a $d\!=\!1\!+\!1$ CFT, and the
isometry group is infinite-dimensional.  The entropy was computed
using older results on the central charge of the CFT and using Cardy's
formula to get the degeneracy of states.  In this approach there were
some technical subtleties, and unresolved issues such as the question
of where the states being counted reside.  The presence of holography
makes localization in the radial coordinate of the AdS factor a tricky
issue.

A lucent discussion of the subtleties in counting the entropy of the
BTZ black hole via different methods, and additional references, may
be found in a recent overview article by Carlip.  We refer the
interested reader to that work \cite{carlip98}.

AdS/CFT methods were extended to rotating $d\!=\!5$ black holes in
\cite{mcrotads}, and the S-dual NS-NS case was studied in 
\cite{mcaat9806}.  The BTZ factor was seen \cite{vbfl98} in the
$d\!=\!4$ case by lifting on $x^\natural$ to M-theory (the other
factor being $S^2$), and $d\!=\!4$ rotation was discussed in
\cite{mcfl9805}.  BPS states were tracked in \cite{exclprin}, and a
nonperturbative stringy exclusion principle was found to be operating.
In that work it was also shown how the discrete identifications giving
BTZ from $AdS_3$ give a natural mechanism for seeing thermal effects
in the black hole background.  The relation between AdS/CFT and
perturbations of black holes was studied in {\em e.g.} \cite{vbpkal}.
Some corrections to leading-order results have been investigated in
\cite{beretal}.

An explanation for the agreement of the D-brane and black hole
entropies for extremal but non-BPS $d\!=\!5$ Type-I black holes has
been offered in \cite{joseetal}.  There it was suggested that
restoration of supersymmetry in the large-$N$ limit is responsible for
the unexpected agreement.  It would be interesting to see whether this
proposal carries over to other known extremal non-BPS black holes.

Use of the AdS/CFT correspondence to learn about black holes via gauge
theory has also been initiated.  In a study of the spectrum of the
large-$N$ gauge theory using supergravity, it was argued in
\cite{gthho} that the gauge theory in the D3 case describes spacetime
behind the horizon as well as the region outside.  In \cite{gthsr},
where nonextremal D3-branes were studied, it was further conjectured
that large-$N$ gauge theory also offers the possibility of resolving
both black hole curvature singularities and the information problem.
An $AdS_3$ map between the boundary CFT and the theory in the bulk 
\cite{ejmmm} was argued to be useful for tracking information in
black hole spacetimes with the BTZ factor.

We look forward to future progress ({\em e.g.} \cite{bdhm}) on black
hole issues using the AdS/CFT correspondence and other aspects of
M/string theory.  Particularly interesting would be insights on the
precise nature and the location of states counted by the
Bekenstein-Hawking entropy, and hints on solving the information
problem.


\end{document}